\documentclass[preprint,showpacs,preprintnumbers,amsmath,amssymb]{revtex4}

\usepackage{graphicx,amsfonts}
\usepackage{epsfig}
\usepackage{dcolumn}
\usepackage{bm}
\begin{document}


\title{Closing the $SU(3)_L\otimes U(1)_X$ Symmetry at Electroweak Scale
}

\author{Alex G. Dias}
\email{alexdias@fma.if.usp.br}
\affiliation{Instituto de F\'\i sica, Universidade de S\~ao Paulo\\
C. P. 66.318, 05315-970 S\~ao Paulo, SP, Brazil.
}
\author{J. C. Montero}
\email{montero@ift.unesp.br}
\affiliation{ Instituto de F\'\i sica Te\'orica,
Universidade Estadual Paulista\\
Rua Pamplona 145, 01405-000, S\~ao Paulo, SP--- Brazil.
}
\author{V. Pleitez}
\email{vicente@ift.unesp.br}
\affiliation{ Instituto de F\'\i sica Te\'orica,
Universidade Estadual Paulista\\
Rua Pamplona 145, 01405-000, S\~ao Paulo, SP--- Brazil.
}
\date{\today}

\begin{abstract}
We show that some models with $SU(3)_C\otimes SU(3)_L\otimes U(1)_X$
gauge symmetry can be realized at the electroweak scale and that
this is a consequence of an approximate global $SU(2)_{L+R}$
symmetry. This symmetry implies a condition among the vacuum
expectation value of one of the neutral Higgs scalars, the
$U(1)_X$'s coupling constant, $g_X$, the sine of the weak mixing
angle $\sin\theta_W$, and the mass of the $W$ boson, $M_W$. In the
limit in which this symmetry is valid it avoids the tree level
mixing of the $Z$ boson of the Standard Model with the extra
$Z^\prime$ boson. We have verified that the oblique $T$ parameter is
within the allowed range indicating that the radiative corrections
that induce such a mixing at the 1-loop level are small. We also
show that a $SU(3)_{L+R}$ custodial symmetry implies that in some of
the models we have to include sterile (singlets of the 3-3-1
symmetry) right-handed neutrinos with Majorana masses, being the
see-saw mechanism mandatory to obtain light active neutrinos.
Moreover, the approximate $SU(2)_{L+R}\subset SU(3)_{L+R}$ symmetry
implies that the extra non-standard particles of these 3-3-1 models
can be considerably lighter than it had been thought before so that
new physics can be really just around the corner.

\end{abstract}

\maketitle

\section{Introduction}
\label{sec:intro}

It is usually assumed that any physics beyond the Standard Model
(SM) must have this model as a good approximation at energies of the
order of hundred GeVs or, in practice, up to the LEP energies. This
implies that the new degrees of freedom should be related to a
sufficiently high energy scale. In particular, in many of the
extensions of the SM there is at least one additional neutral gauge
boson, generically denoted by $Z^\prime$, whose mass has to be of
the order of few TeV in order to keep consistency with present
phenomenology. This is the case, for instance, of the left-right
models~\cite{lr}, models with an extra $U(1)$
factor~\cite{aguila,carena,babu,pdg}, and in grand unified theories
with symmetries larger than $SU(5)$ as $SO(10)$ and
$E_6$~\cite{hewett}, little Higgs scenarios~\cite{lh}, and models
with extra dimensions~\cite{kk}. In all these cases, the existence
of additional real neutral vector bosons yields deviations from the
condition $M_W=\cos\theta_WM_Z$ because of the mixing between the
$Z$-boson of the  SM and the new neutral vector boson  $Z^\prime$.
There are deviations also in the neutral current parameters of the
fermion $i$ in the vector and axial-vector interactions with $Z$,
denoted by $g^i_{V,A}$. These parameters only coincide within
certain approximation with those of the $Z$. In general these
deviations are proportional to $(v_W/\Lambda)^2$ or higher power of
this ratio, where $\Lambda$ is an energy scale, say a vacuum
expectation value (VEV), related to the breaking of the hidden extra
symmetry. So, it is thought to be necessary that $\Lambda\gg
v_W\approx 246$ GeV in order to make the models compatible with the
present phe\-no\-me\-no\-lo\-gy. This makes the search for extra
neutral gauge bosons one of the main goals of the next high energy
collider experiments~\cite{lhcilc}.

Usually, the interactions involving $Z^\prime$  are parameterized
(besides the pure kinetic term) as~\cite{babu,pdg}
\begin{eqnarray}
{\mathcal  L}^{NC(Z^\prime)}&=&-
\frac{\sin\xi}{2}F^\prime_{\mu\nu}F^{\mu\nu}+M^2_{Z^\prime}Z^\prime_\mu
Z^{\prime\mu}+\delta M^2Z^\prime_\mu Z^{\mu}\nonumber \\ &-&
\frac{g}{2 c_W}
\sum_i\bar{\psi}_i\gamma^\mu(f^i_V-f^i_A\gamma^5)\psi_i Z^\prime_\mu,
\label{zgeral}
\end{eqnarray}
where $Z$, which is the would be neutral vector boson of the SM,
and $Z^\prime$ are not yet mass eigenstates, having a mixing defined by
the angle
\begin{equation}
\tan 2\phi=\frac{\delta M^2}{M^2_{Z^\prime}-M^2_Z},\label{phi}
\end{equation}
and where $c_W\equiv \cos \theta_W$ (and for future use $s_W\equiv
\sin \theta_W$). If $Z_1$ and $Z_2$ denote the mass eigenstates,
then in most of the models we have $M_{Z_2} \gg M_{Z_1} \approx
M_Z$, and hence $\phi\ll1$. In this situation the  vector and the
axial-vector neutral current parameters, $g^{i(SM)}_V$ and
$g^{i(SM)}_A$, respectively, of the $Z$-boson with the known
fermions are shifted, at tree level, as follows:

\begin{equation}
g^i_V= g^{i(SM)}_Vc_\phi+ f^i_V s_\phi,\quad\quad g^i_A= g^{i(SM)}_Ac_\phi+
 f^i_As_\phi,
\label{shift}
\end{equation}
where $g^{i(SM)}_V\!\!=\!\!T^i_3-2Q_i s_W^2$ and
$g^{i(SM)}_A=T^i_3$, being $T^i_3=\pm1/2$ and $Q_i$ the electric
charge of the fermion $i$; we have used the notation
$c_\phi\,(s_\phi)=\cos\phi\,(\sin\phi)$.  The parameters $f^i_{V,A}$
in Eq.~(\ref{shift}) are not in general the same for all particles
of the same electric charge, thus, we have flavor changing neutral
currents (FCNC) coupled to $Z$ which imply strong constraints coming
from experimental data such as $\Delta M_K$ and other $\vert\Delta
S\vert=2$ processes. These constraints imply a small value for the
mixing angle $\phi$ or, similarly, a large value for the energy
scale  $\Lambda$. If $s_\phi=0$ is imposed such constraints could be
avoided, however in most of the models with $Z^\prime$ this usually
implies a fine tuning among $U(1)$ charges and vacuum expectation
values that is far from being natural~\cite{carena}.

In principle, a shift as in Eq.~(\ref{shift}) occurs in 3-3-1
models~\cite{331,pt,mpp}. These models have a rich scalar sector
that implies, in general, a mixing of $Z$, the vector boson of
$SU(2)_L\subset SU(3)_L$, and $Z^\prime$, the gauge boson related to
the $SU(3)_L$ symmetry. Working in the $Z,Z^\prime$ basis the
condition $\sin\phi\ll1$ can be obtained if the energy scale
$\Lambda$ (which in these models is identified with the VEV that
breaks the $SU(3)_L$ symmetry, $v_\chi$) is above the TeV scale.
Hence, it is usually believed that only approximately we can have
that $Z_1\approx Z$, even at the tree level. The same happens with
the neutral current parameters, $g^{i}_{V,A}$, which only
approximately coincide with $g^{i(SM)}_{V,A}$. This is true since
the corrections to the $Z$ mass and $g^{i}_{V,A}$ in these models,
assuming $v_\chi\gg v_W \simeq 246$ GeV, are proportional to
$(v_W/v_\chi) ^{2}$ and for $v_\chi\to\infty$ we recover exactly the
SM with all its degrees of freedom, with the heavier ones introduced
by the $SU(3)_L$ symmetry decoupled. However, we expect that
$v_\chi$ should not be extremely large if new physics is predicted
to show up in the near future experiments. In practice, measurements
of the $\rho_0$ parameter, and FCNC processes like $\Delta M_K$,
should impose constraints upon the $v_\chi$ scale at which the
$SU(3)_L$ symmetry arises.

However, it was pointed out recently in Ref.~\cite{newpca} that in
3-3-1 models, at the tree level, it is possible that: i) there is no
mixing between $Z$ and $Z^\prime$, and the latter boson may have a
mass even below the TeV scale; ii) $\rho_0=1$ since $M_{Z_1}=M_Z$,
and iii) the vector and axial-vector parameters in the neutral
currents coupled to $Z_1$, $g^i_{V,A}$, being exactly those of the
SM, $g^{i(SM)}_{V,A}$, independently of the $v_\chi$ value. This is
implied not by a fine tuning but by a condition which can be
verified experimentally involving the parameters of the model, $g$,
$M_W$, $s_W$ and one of the VEVs. Such condition is a consequence of
an approximate global $SU(2)_{L+R}$ symmetry. In the limit of the
exact symmetry we have $\sin\phi=0$, avoiding in this way the shift
as in Eq.~(\ref{shift}), and $\sin\theta_W=0$ as in the SM.
Remarkably, when $\sin\phi=0$ but $\sin\theta_W\not=0$ the
parameters in the neutral currents coupled to the heavy boson $Z_2$
depend only on the weak mixing angle $\theta_W$, meaning that they
are not free parameters anymore. Moreover the couplings of $Z_2$
with leptons are suppressed by the leptophobic factor
$(1-4\sin^2\theta_W)^{1/2}$ ~\cite{dumm97}.

The outline of this paper is as follows. In Sec.~\ref{sec:model} we
review briefly the three 3-3-1 models that will be considered here.
We give the representation content of the model with bileptons
(Sec.~\ref{subsec:mm}), with heavy leptons (Sec.~\ref{subsec:hl})
and with right-handed neutrinos in (Sec.~\ref{subsec:rn}). Next, in
Sec.~\ref{sec:gb} we give exact expressions for the gauge vector
boson eigenstates and their respective masses for the three models.
In Sec.~\ref{subsec:smmm}--\ref{subsec:smnr} we give what we call
the ``SM limit" for each model. In Sec.~\ref{sec:nc} we show the
exact expressions for the parameters $g^i_{V,A},f^i_{V,A}$ appearing
in the neutral currents coupled to $Z_1$ and $Z_2$:
Sec.~\ref{subsec:ncmm} for the case of the model with bileptons;
Sec.~\ref{subsec:nchl} for the heavy lepton models and,
Sec.~\ref{subsec:ncnr} for the model with right-handed neutrinos. We
also show in Secs.~\ref{subsec:ncmm}-\ref{subsec:ncnr} that if we
impose that there is no mixing between $Z$ and $Z^\prime$, then
$g^i_{V,A}$ coincide exactly with the respective parameters of the
SM's $Z$ boson for all the known particles. It means that there is
no flavor changing neutral current in the known sector and that
$f^i_{V,A}$ depend only on $s_W$. In Sec.~\ref{sec:custodial} we
explain the small value for $\phi$ as consequence of a global
approximate custodial symmetry. These results are interpreted in the
last section, Sec.~\ref{sec:con}. In the appendix we show that there
is also an approximate global $SU(3)_{L+R}$ symmetry which, although
badly broken, is useful for obtaining a realistic mass spectra in
the scalar sectors. Moreover we also discuss that this extended
custodial symmetry implies that it is mandatory to include
right-handed sterile (with respect to the 3-3-1 symmetry) neutrinos
and to consider the see-saw mechanism to obtain light active
neutrinos.

\section{The models}
\label{sec:model}

Models with $SU(3)_C\otimes SU(3)_L\otimes U(1)_X$ gauge symmetry
(called 3-3-1 models for short) are interesting  possibilities for
the electroweak interactions at the TeV
scale~\cite{331,pt,mpp,singer}. At low energies it is expected that,
like any other extension of the SM, they must coincide with this
model. By choosing appropriately the representation content they
give at least partial explanations to some fundamental questions
that are accommodated but not explained by the SM~\cite{cp6}.

The 3-3-1 models which embed the Standard Model are those of
Refs.~\cite{331,pt,mpp,singer}. Here we will consider only: 1) the
minimal model which has charged bileptons \cite{331}, 2) the model
with heavy leptons \cite{pt}, and 3) the model with right-handed
neutrinos~\cite{mpp}. These models have a electric charge operator
that can be written as
\begin{equation}
\frac{Q}{e}=T_3-b\,T_8+X,
\label{co}
\end{equation}
where $T_i,\,i=3,8$ are the diagonal generators of $SU(3)$ and
$b=\sqrt{3}$ for the minimal model~\cite{331} and also for the model
with heavy leptons ~\cite{pt}, while $b=1/\sqrt3$ for the model with
right-handed neutrinos transforming non-trivially under
$SU(3)_L$~\cite{mpp} or heavy neutral leptons~\cite{singer}.

\subsection{The minimal model}
\label{subsec:mm}

Let us consider first the minimal 3-3-1 model~\cite{331} in which
$b=\sqrt3$ and the known leptons transform as triplets:
\begin{eqnarray}
\Psi_{aL}=(\nu_a,\,l_a,\,l^c_a)^T_L \sim({\bf1},{\bf3},0), \quad
\nu_{aR}\sim({\bf1},{\bf1},0),
\label{lep1}
\end{eqnarray}
here $T$ means transpose and $a=e,\mu,\tau$.

In the quark sector we have two anti-triplets and one triplet:
\begin{eqnarray}
Q_{mL}=(d_m,\, u_m,\, j_m)^T_L\sim({\bf3}, {\bf3}^{*},- 1/3),
\hspace{.7 cm}
Q_{3L}=(u_3,\, d_3,\,J)^T_L\sim({\bf3}, {\bf 3}, 2/3),
\label{q1l}
\end{eqnarray}
with $m=1,2$, and the right-handed components transforming as
singlets:
\begin{eqnarray}
& & u_{\alpha R}\sim({\bf3},{\bf1},2/3), \quad
d_{\alpha R} \sim({\bf3},{\bf1},-1/3), \;\alpha=1,2,3;\nonumber\\
& & J_{R}\sim({\bf3},{\bf1},5/3), \quad
j_{mR}\sim({\bf3},{\bf1},-4/3).
\label{q1r}
\end{eqnarray}

The minimal scalar sector of the model consists of three triplets:
\begin{eqnarray}
& &\eta=(\eta^0,\,\eta^{-}_1,\,\eta^+_2)^T\sim({\bf1},{\bf3},0),\nonumber\\
& &\rho=(\rho^+,\,\rho^0,\,\rho^{++})^T\sim({\bf1},{\bf3},1), \nonumber\\
& &\chi=(\chi^-,\,\chi^{--},\,\chi^0)^T\sim({\bf1},{\bf3},-1).
\label{trim1}
\end{eqnarray}
and the sextet
\begin{equation}
S=\left(\begin{array}{ccc}
\sigma^0_1 & \frac{h_2^+}{\sqrt2}\; & \;\frac{h_1^-}{\sqrt2} \\
\frac{h_2^+}{\sqrt2} & H_1^{++}\; &\; \frac{\sigma_2^0}{\sqrt2} \\
\frac{h_1^-}{\sqrt2} &\frac{\sigma_2^0}{\sqrt2}\; & \; H_2^{--}
\end{array}\right)\sim({\bf6}^*,0).
\label{sextet}
\end{equation}

The VEVs for the scalar Higgs multiplets are denoted by
$\langle\eta^0\rangle=v_\eta/\sqrt2$,
$\langle\rho\rangle=v_\rho/\sqrt2$,
$\langle\chi^0\rangle=v_\chi/\sqrt2$, for the triplets, and
$\langle\sigma^0_2\rangle=v_s$ and $\langle\sigma^0_1\rangle=0$, in
the scalar sextet, i.e., we are neglecting left-handed neutrino
Majorana masses.

Since the extra quarks have all exotic electric charges, the mixing
in the known quark sectors are exactly as in the SM. In the neutrino
sector, the presence of sterile neutrinos allows a general mass
matrix with both, Dirac and Majorana masses.

\subsection{The model with heavy leptons}
\label{subsec:hl}

In the model of Ref.~\cite{pt}, which also has $b=\sqrt3$, it is
introduced, in each lepton triplet, a heavy charged field $E^+$:
\begin{eqnarray}
\Psi_{aL}=(\nu_a,\,l^-_a,\,E^+_a)^T_L \sim({\bf1},{\bf3},0),
\label{lep2l}
\end{eqnarray}
and the right-handed components of the leptons transforming as
\begin{eqnarray}
\nu_{aR} \sim({\bf1},{\bf1},0); \quad l^-_{aR}
\sim({\bf1},{\bf1},-1); \quad E^+_{aR} \sim({\bf1},{\bf1},1).
\label{lep2r}
\end{eqnarray}

The quark sector is the same of the previous model and here only the
triplets in Eq.~(\ref{trim1}) are needed for given to fermions and
gauge bosons appropriate masses.

The mixing in the quark sector is as in the previous model. However,
in the charged lepton sector it is possible to have a general mixing
between the known charged leptons, $l^-_a$, and the heavy ones,
$E^-_a$. For instance, interactions such as
$\epsilon\overline{(\Psi_{aL})^c}\Psi_{bL}\eta$ and
$\overline{l^c_{aL}}E_{bR}$ induce such a mixture. This can be
avoided by introducing an appropriate discrete symmetry. On the
other hand, neutrinos have Dirac masses and the right-handed sterile
ones can get a Majorana mass term. Although in this model the scalar
sextet is not necessary, it can be introduced to generate Majorana
mass terms for active left-handed neutrinos.

\subsection{Model with right-handed neutrinos}
\label{subsec:rn}

If right-handed sterile neutrinos do exist then it is possible that
they transform non-trivially under a larger gauge symmetry group,
for instance the 3-3-1 symmetry with $b=~1/\sqrt3$~\cite{mpp}. This
model is probably the more economical one to incorporate sterile
neutrinos with respect to the SM interactions~\cite{alexpires}.

In this case the representation content is as follows~\cite{mpp}:
\begin{eqnarray}
\psi_{aL}=(\nu_a,\,e_a,\,\nu^c_a)^T_L \sim({\bf1},{\bf3},-1/3),
\label{lep3l}
\end{eqnarray}
and the right-handed components for the charged leptons,
\begin{eqnarray}
e_{aR} \sim({\bf1},{\bf1},-1).
\label{lep3r}
\end{eqnarray}

The quark sector consists in the following representations:
\begin{eqnarray}
Q_{mL}=(d_m,\, u_m,\, D_m)^T_L\sim({\bf3}, {\bf3}^{*},0),
\hspace{.7 cm}
Q_{3L}=(u_3,\, d_3,\,U)^T_L\sim({\bf3}, {\bf 3}, 1/3),
\label{q2l}
\end{eqnarray}
with $m=1,2$. And the respective right-handed components:
\begin{eqnarray}
& & u_{ \alpha R}\sim({\bf3},{\bf1},2/3), \hspace{.5 cm}
d_{\alpha  R} \sim({\bf3},{\bf1},-1/3), \nonumber\\
& & U_R \sim({\bf3},{\bf1},2/3) , \hspace{.5 cm}
D_{m R} \sim({\bf3},{\bf1},-1/3),
\label{q2r}
\end{eqnarray}
with $\alpha =1,2,3$.

The scalar sector of the model is:
\begin{eqnarray}
& &\eta=(\eta^0,\,\eta^{-},\,\eta^{ \prime0})^T\sim({\bf1},{\bf3},-1/3),
\nonumber\\
& &\rho=(\rho^+,\,\rho^0,\,\rho^{\prime+})^T\sim({\bf1},{\bf3},2/3),
\nonumber\\
& &\chi=(\chi^0,\,\chi^{-},\,\chi^{ \prime0})^T\sim({\bf1},{\bf3},-1/3).
\label{t2}
\end{eqnarray}

The only nonzero VEVs are $\langle\eta^0\rangle=v_\eta/\sqrt2$,
$\langle\rho^0\rangle=v_\rho/\sqrt2$ and
$\langle\chi^{\prime0}\rangle=v_{\chi^{\prime}}/\sqrt2$.

In order to avoid favor changing neutral currents (FCNC) this model
has been considered with three scalar triplets of the sort showed in
Eq.~(\ref{t2}), see Refs.~\cite{mpp}. Only two of them are necessary
to give mass to all fermions and to implement the spontaneous
breaking of the gauge symmetry~\cite{dong06}. A version of the model
with four scalar triplets generates fermion masses without a
hierarchy in the Yukawa couplings~\cite{carlos}. However, in this
case the equivalent to the Cabibbo-Kobayashi-Maskawa mixing matrix
is not unitary and there is also FCNC mediated by the Z vector boson
of the SM. In general depending on: the number of scalar triplets,
on discrete symmetries and on the VEV structure, the model could
have, or not, FCNC in all charged sectors, and the mixing matrix in
the couplings with $W^\pm$ could be, or not, exactly the same as in
the SM. An scalar sextet can also be introduced in order to have
more space to generate neutrino masses~\cite{singer,kyvan}.

\section{Gauge boson masses and eigenstates}
\label{sec:gb}

From the  kinetic terms for the scalar fields, constructed with the covariant
derivatives
\begin{eqnarray}
{\mathcal D}_\mu\varphi=\partial_\mu\varphi - ig {\cal M}_\mu \varphi
-ig_{_X} X\varphi B_\mu,\nonumber
\label{dct}
\end{eqnarray}
\begin{equation}
{\mathcal D}_\mu S = \partial_\mu S -ig \left[ {\cal M}_\mu   S +
S^T {\cal M}^T_\mu     \right],
\label{dcs}
\end{equation}
where $g_{_X}$ denotes the U(1)$_X$ gauge coupling constant and
$\varphi=\eta,\rho,\chi$, we can obtain the mass matrices for the
vector bosons.

Defining ${\cal M}_\mu\equiv \vec{W}_\mu\cdot \vec{T}$, we have
\begin{eqnarray}
{\cal{M}}_\mu= \left(\begin{array}{ccc}
  W^3_\mu + \frac{1}{\sqrt 3}W^8_\mu+2t X_\varphi B_\mu & \sqrt{2}W_\mu^{+} &
  \sqrt{2}(V_\mu^{\frac{1}{2}(\sqrt3 b-1) })^*\\
  \sqrt{2}W_\mu^{-} & -W^3_\mu+ \frac{1}{\sqrt 3}W^8_\mu+2t   X_\varphi B_\mu &
  \sqrt{2}(U_\mu^{\frac{1}{2}(\sqrt3 b+1) })^* \\
 \sqrt{2}V_\mu^ {\frac{1}{2}(\sqrt3 b-1) } & \sqrt{2}U_\mu^ {\frac{1}{2}(\sqrt3
 b+1) } & -\frac{2}{\sqrt 3}W^8_\mu+2t  X_\varphi B_\mu
\end{array}\right),
\label{mbv}
\end{eqnarray}
with $t\equiv{g_X}/{g}$ and where the non hermitian gauge bosons are defined as
\begin{eqnarray}
W_\mu^{+}=(W^1_\mu-iW^2_\mu)/\sqrt 2,\nonumber \\
V_\mu^{\frac{1}{2}(\sqrt3 b-1) }=(W^4_\mu+iW^5_\mu)/\sqrt 2, \label{wvu}\\
U_\mu^{\frac{1}{2}(\sqrt3 b+1) }=(W^6_\mu+iW^7_\mu)/\sqrt 2,\nonumber
\label{defbc}
\end{eqnarray}
with $ \pm {\frac{1}{2}(\sqrt3 b\pm 1) }$ denoting the electric
charge in units of the $\vert e \vert$ of the heavy gauge bosons $V$
and $U$. In 3-3-1 models with $b=\sqrt3$ both heavy vector bosons
are charged, $V^\pm$ and $U^{\pm\pm}$. In 3-3-1 models with
$b=1/\sqrt3$ we have $U^\pm$ and a non-hermitian $V^0$ neutral
vector bosons.

In the minimal model, the mass square of the non-hermitian vector
bosons are given by
\begin{equation}
M^2_W=\frac{1}{4}\,g^2 v^2_{_W}, \,\,
M^2_V=\frac{1}{4}\,g^2\left(v^2_\eta+2v^2_s+v^2_\chi\right), \,\,
M^2_U=\frac{1}{4}\,g^2\left(v^2_\rho+2v^2_s+v^2_\chi\right),
\label{mbc}
\end{equation}
where $v^2_{_W}\equiv v^2_\eta+v^2_\rho+2v^2_s$. Notice that as
$v_{_W}\approx246$ GeV, the usual VEV of the Higgs in the SM, then
$v_\chi$ must be, in principle, large enough in order to keep the
new gauge bosons, as $V$ and $U$, sufficiently heavy to be
consistent with the present experimental data. In models where the
sextet is not necessary those expressions in Eq.~(\ref{mbc}) are
still valid puting $v_s=0$. In all these models there is no mixing
between $Z$ and $Z^\prime$ in the kinetic term at the tree level,
thus $\sin\xi=0$ in Eq.~(\ref{zgeral}).

Insofar the analysis is valid for both $b=\sqrt{3}$ and $b=1/\sqrt{3}$
models, however in order to clarify, when considering the real neutral
gauge bosons we will study both cases separately.

\subsection{Neutral gauge bosons in the minimal 3-3-1 model}
\label{subsec:gbmmhl}

The mass matrix for the real neutral vector bosons in the
$(W^3_\mu $, $W^8_\mu $,  $B_\mu)$ basis is
\begin{eqnarray}
M^2_{(b=\sqrt{3})}=\frac{g^2}{4}v_\chi^2\left(\begin{array}{ccc}
\overline{v}_{_W}^2 & \frac{1}{\sqrt 3}(
\overline{v}_{_W}^2-2\overline{v}_\rho^2 ) & -2t \overline{v}_\rho^2  \\
\frac{1}{\sqrt 3}( \overline{v}_{_W}^2-2\overline{v}_\rho^2 ) &
\frac{1}{3}(\overline{v}_{_W}^2 + 4)&
 \frac{2}{\sqrt 3}t(\overline{v}_\rho^2 +2) \\
-2t \overline{v}_\rho^2 & \frac{2}{\sqrt 3}t( \overline{v}_\rho^2 +2) &
4t^2(\overline{v}_\rho^2 +1)\end{array}\right),
\label{mnt}
\end{eqnarray}
and we have introduced the dimensionless ratios
$\overline{v}_\rho={v_\rho}/{v_\chi}$ and $\overline{v}_{_W}={v_W}/{v_\chi}$.
The matrix in Eq.~(\ref{mnt}) has a vanishing eigenvalue corresponding to the
photon. The other two ei\-gen\-va\-lues, $M_{Z_1}$ and $M_{Z_2}$, can be written
exactly, introducing two dimensionless parameters $m_1$
and $m_2$, as
\begin{equation}
m^2_1 \equiv \frac{2M_{Z_1}^2}{g^2v_\chi^2}=A(1-R);
\quad m^2_2\equiv \frac{2M^2_{Z_2}}{g^2v_\chi^2}=A(1+R),
\label{mz1z2mm}
\end{equation}
where we have defined
\begin{equation}
A=\frac{1}{3}\left[3t^2(\overline{v}_\rho^2 +1 )
+\overline{v}_{_W}^2+1\right],
\label{amm}
\end{equation}
and
\begin{equation}
R=\left[1-\frac{1}{3A^2}\,(4t^2+1)[
\overline{v}^2_{_W}(\overline{v}_\rho^2
+1)-\overline{v}_\rho^4]\right]^{\frac{1}{2}},
\label{rmm}
\end{equation}
with $t$ given by $t^2=s^2_W/(1-4s^2_W)$ (see below).
Notice that instead of introducing a mixing angle between the $Z$ and the
$Z^\prime$ bosons, as in Refs.~\cite{ng}, we have diagonalized directly the
mass square matrix in Eq.~(\ref{mnt}).

The eigenstates of the symmetry $W^3_\mu$, $W^8_\mu $ and $ B_\mu$
can be written in terms of the mass eigenstates $A_\mu,Z_{1\mu}$ and
$Z_{2\mu}$ in an exact form as
\begin{eqnarray}
&W^3_\mu&=  \frac{ t}{\sqrt{4t^2+1}}A_\mu-N_1\left( 3 m^2_2 +
\overline{v}_\rho^2 -2\overline{v}_{_W}
^2\right)Z_{1\mu}-N_2\left(3m^2_1 +  \overline{v}_\rho^2
-2\overline{v}_{_W} ^2\right)Z_{2\mu},
 \nonumber \\
& \frac{ W^8_\mu} {\sqrt3} &= -  \frac{ t}{\sqrt{4t^2+1}}A_\mu -  N_1\left(
m^2_2 +  \overline{v}_\rho^2 -\frac{2}{3}\overline{v}_{_W} ^2-
\frac{2}{3}   \right)Z_{1\mu}- N_2\left(m^2_1 +  \overline{v}_\rho^2
-\frac{2}{3} \overline{v}_{_W} ^2-\frac{2}{3}   \right)Z_{2\mu},
\nonumber \\
&B_\mu&=  \frac{ 1}{\sqrt{4t^2+1}}A_\mu+2t (1- \overline{v}_\rho^2 )N_1
Z_{1\mu}+2t(1-\overline{v}_\rho^2 )N_2 Z_{2\mu} ,
\label{w3w8b}
\end{eqnarray}
with the normalization factors
\begin{eqnarray}
  & &N^{-2}_1 = 3  \left(2m^2_2  +\overline{v}_\rho^2
  -\frac{4}{3}\overline{v}_{_W} ^2-\frac{1}{3} \right)^2 + ( \overline{v}_\rho^2
  -1)^2(4t^2+1),
 \nonumber\\
& &N^{-2}_2= 3  \left(2m^2_1  +\overline{v}_\rho^2
-\frac{4}{3}\overline{v}_{_W} ^2-\frac{1}{3} \right)^2 + ( \overline{v}_\rho^2
-1)^2(4t^2+1).
\label{norm}
\end{eqnarray}

Only the components in $A_\mu$ do not depend on the VEVs
but the others in $Z_{1,2}$ do. The interaction of the photon with leptons is
therefore
\begin{eqnarray}
g \frac{ t}{\sqrt{4t^2+1}}\, \overline{l_a}\gamma^\mu l_aA_\mu =
e\,\overline{l_a}\gamma^\mu l_aA_\mu = g
\,s_W\,\overline{l_a}\gamma^\mu l_aA_\mu,
\label{etsw}
\end{eqnarray}
where $e$ is the electric charge of the positron.
We can identify $e=gs_W$ since in 3-3-1 models $SU(2)_L\subset SU(3)_L$, i. e.,
$g_{_{SU(3)_L}} \equiv g_{_{SU(2)_L}}$. On the other hand, the condition
$1/e^2=4/g^2+1/g^2_X$ for 3-3-1 models with
$b=\sqrt3$~\cite{331} implies

\begin{equation}
t^2\equiv \frac{\alpha_X}{\alpha_L}= \frac{s^2_W}{1-4s^2_W},
\label{polo}
\end{equation}
with $\alpha_i=g_i^2/4\pi$, $i=X,L$, where we have introduced the
notation $g\equiv g_L$. Notice, for future use in
Sec~\ref{sec:custodial} and in the Appendix, that $s_W=0$ implies
$g_X=0$.

\subsection{Neutral gauge bosons in the model with right-handed neutrinos}
\label{subsec:gbmnr}

The masses of the charged vector bosons are as in Eqs.~(\ref{mbc}) but
with $v_s=0$. In this case the mass square matrix for the real neutral bosons
in the $(W^3_\mu $,  $W^8_\mu $,  $B_\mu)$ basis is
\begin{eqnarray}
M^2_{(b=1/\sqrt{3})}=\frac{g^2}{2}v_\chi^2\left(\begin{array}{ccc}
\frac{1}{2} \overline{v}_{_W}^2 & \frac{1}{2\sqrt 3}(
\overline{v}_{_W}^2-2\overline{v}_\rho^2 ) & -\frac{1}{3}t ( \overline{v}_{_W}^2
+\overline{v}_\rho^2 ) \\
\frac{1}{2\sqrt 3}( \overline{v}_{_W}^2-2\overline{v}_\rho^2 ) &
\frac{1}{6}(\overline{v}_{_W}^2 + 4)&
 \frac{1}{3\sqrt 3}t(3\overline{v}_\rho^2 -\overline{v}_{_W}^2+2) \\
-\frac{1}{3}t ( \overline{v}_{_W}^2 +\overline{v}_\rho^2 ) &  \frac{1}{3\sqrt
3}t(3\overline{v}_\rho^2 -\overline{v}_{_W}^2+2) &
\frac{2}{9}t^2(3\overline{v}_\rho^2+\overline{v}_{_W}^2 +1)\end{array}\right).
\label{mntnd}
\end{eqnarray}

As in the previous section we define the dimensionless parameters
for this model $m^2_1$ and $m^2_2$ as in Eq~(\ref{mz1z2mm}) but now
the VEV appearing in them is $v_{\chi^\prime}$, and $A$ and $R$ are
given by
\begin{equation}
A=\frac{1}{9}\,\left[t^2(  \overline{v}_{_W} ^2+3\overline{v}_\rho^2
+1 ) +3(\overline{v}_{_W} ^2+1)\right], \label{ap}
\end{equation}
and
\begin{equation}
R=\left[1-\frac{1}{9A^{ 2}}\,(4t^2+3)
[\overline{v}^2_{_W}(\overline{v}^2_\rho+1)-\overline{v}^4_\rho]\right]^{\frac{1}{2}}.
\label{rp}
\end{equation}

The symmetry eigenstates $W^3_\mu$, $W^8_\mu $ and $ B_\mu$ can be
written in terms of the mass eigenstates $A_\mu$, $Z_{1\mu} $ and
$Z_{2\mu}$:
\begin{eqnarray}
&W^3_\mu&=  \frac{ \sqrt{3}t}{ \sqrt{4t^2+3}}A_\mu-N_1\left[ 3
m^2_2
-3( \overline{v}_{_W} ^2- \overline{v}_\rho^2 ) \right]Z_{1\mu}-
N_2\left[3
m^2_1 -3( \overline{v}_{_W} ^2- \overline{v}_\rho^2 ) \right]Z_{2\mu},
 \nonumber \\ \nonumber \\
& \frac{ W^8_\mu} {\sqrt3} &= -  \frac{ 1}{3}  \frac{  \sqrt{3}t}
{\sqrt{4t^2+3}}A_\mu -  N_1\left[ 3 m^2_2 -3( \overline{v}_{_W} ^2-
\overline{v}_\rho^2 ) \right]Z_{1\mu}- N_2\left[3 m^2_1 -3(
\overline{v}_{_W} ^2- \overline{v}_\rho^2 ) \right]Z_{2\mu},
\nonumber \\ \nonumber\\
&B_\mu&=  \frac{ \sqrt{3}}{ \sqrt{4t^2+3}}A_\mu+2t (1+
\overline{v}_\rho^2-\overline{v}_{_W} ^2 )N_1
Z_{1\mu}+2t(1+\overline{v}_\rho^2-\overline{v}_{_W} ^2 )N_2 Z_{2\mu},
\label{w3w8bnd}
\end{eqnarray}
with the normalization factors
\begin{eqnarray}
  & &N^{-2}_1 = 9  \left(2m^2_2  +\overline{v}_\rho^2
  -\overline{v}_{_W} ^2-1 \right)^2 + ( 1+\overline{v}_\rho^2 - {v}_{_W}
  ^2)^2(4t^2+3),
  \nonumber\\
& &N^{-2}_2= 9  \left(2m^2_1  +\overline{v}_\rho^2 -
\overline{v}_{_W}^2-
1 \right)^2 + ( 1+\overline{v}_\rho^2 - {v}_{_W} ^2)^2(4t^2+3).
\label{normnd}
\end{eqnarray}
As in the previous case only the components on $A_\mu$ do not depend on the
VEVs. In this sort of models we have
\begin{equation}
t^2= \frac{\alpha_X}{\alpha_L}= \frac{s^2_W}{1-\frac{4}{3}s^2_W}.
\label{polond}
\end{equation}
As in the previous models $s_W=0$ implies $g_X=0$.

\section{SM limit from 3-3-1 models}
\label{sec:sml}

Let us introduce the parameters $\rho_0=c^2_WM^2_Z/M^2_W$ and
$\rho_1$ defined, at the tree level, as
\begin{equation}
\rho_1 \equiv \frac{c^2_W\,M^2_{Z_1}}{M^2_W} =
\frac{2c^2_W}{\overline{v}_{_W} ^2}\,A(1-R), \label{rho1}
\end{equation}
where $A$ and $R$ are defined in Eqs.~(\ref{amm}) and (\ref{rmm}),
respectively. Notice that we are using the inverse of the Standard
Model $\rho_0$ definition~\cite{pdg}. At the tree level $\rho_0=1$
is a prediction of the SM. Thus we call the SM limit of the 3-3-1
model the condition $\rho_1=1$. It means that we are looking for a
relation among the parameters of this model such that $\rho_1=1$. We
have verified that in general $\rho_1\leq1$ and that $\rho_1=1$ is
obtained only under two conditions: \textit{i)} when
$v_\chi\to\infty$, in practice when it is very large, as we have
mentioned in the Introduction; or \textit{ii)} the less trivial
condition when $\overline{v}_\rho$ has a particular value that we
denote $\tilde{v}_\rho$.

On the other hand, we saw that the mass eigenstates $A_\mu,
Z_{1\mu}$ and $Z_{2\mu}$ obtained by inverting the Eq.~(\ref{w3w8b})
have a complex structure that depends on the VEVs for the cases of
$Z_{1\mu}$ and $Z_{2\mu}$. It means that $Z_{1\mu}$ and $Z_{2\mu}$
have, in general, couplings with fermions that are also functions of
the VEVs, of the electric charges and of the weak mixing angle
$\theta_W$. In the so called SM limit, we will obtain that neither
$Z_{1\mu}$ nor $Z_{2\mu}$ depend on the VEVs and, consequently, the
neutral current parameters of $Z_{1\mu}$ are the same as those of
the SM for all known fermions.

\subsection{SM limit in the minimal model}
\label{subsec:smmm}

For the minimal model using $A$ and $R$ given by Eqs.~(\ref{amm})
and (\ref{rmm}), respectively, the non-trivial solution for
obtaining $\rho_1=1$ in Eq.~(\ref{rho1}) is
\begin{eqnarray}
\tilde{\overline{v}}_\rho^2= \frac{1-4s^2_W}{2c^2_W}\,
\overline{v}^2_{_W},
\label{sol3}
\end{eqnarray}
and, since $v_i$, the VEVs of the scalars transforming as doublets under
$SU(2)$, must satisfy the condition $\sum_i v^2_i=(246\,\textrm{GeV})^2$,
Eq.~(\ref{sol3}) also implies
\begin{equation}
\tilde{\overline{v}}^2_\eta+2\tilde{\overline{v}}^2_s=
\frac{1+2s^2_W}{2c^2_W}\,\overline{v}^2_W.
\label{con1}
\end{equation}
Notice that these relations do not depend on the $v_\chi$ scale.
With the present value of $s^2_W=0.2312$~\cite{pdg} we obtain from
Eq.~(\ref{sol3}) and (\ref{con1}), respectively, that
$\tilde{v}_\rho\approx54$ GeV and
$\sqrt{\tilde{v}^2_\eta+2\tilde{v}^2_s}\approx240$ GeV.

Using Eq.~(\ref{sol3}) in the exact eigenvalues of Eq.~(\ref{w3w8b})
we obtain \textit{exactly}, in the basis $(W^3,W^8,B)$, the mass
eigenstates
\begin{eqnarray}
&&\tilde{A}_\mu=(s_W,-\sqrt{3}s_W,\sqrt{1-4s^2_W})\equiv A_\mu,\quad
\tilde{Z}_{1\mu}=(-c_W,-\sqrt{3}\,t_Ws_W,\sqrt{1-4s^2_W}\,t_W)\equiv Z_\mu,\nonumber \\ &&
\tilde{Z}_{2\mu}=(0,\sqrt{1-4s^2_W}/c_W,\sqrt{3}\,t_W)\equiv Z^\prime_\mu,
\label{az1z2}
\end{eqnarray}
where $t_W=s_W/c_W$. Hereafter, the tilde in a quantity $x$ i.e., $\tilde{x}$,
indicates that we are using Eq.~(\ref{sol3}) in the exact expression of $x$.
If we substitute the condition in Eq.~(\ref{sol3}) and (\ref{con1}) in
Eqs.~(\ref{mz1z2mm}) we obtain
\begin{eqnarray}
& & \tilde{M}^2_{Z_1}=\frac{g^2}{4c^2_W}\;v^2_{_W}\equiv M^2_Z,\nonumber \\
& & \tilde{M}^2_{Z_2}=\frac{g^2}{2}\frac{(1-2s^2_W)(4+
\overline{v}^2_{_W})+s^4_W(4-\overline{v}^4_{_W})}{6c^2_W(1-4s^2_W)}\;v^2_\chi
\equiv M^2_{Z^\prime}.
\label{oba1}
\end{eqnarray}

Moreover, the mass of $Z_2$ can be large even if $v_\chi$
is of the order of the electroweak scale. In fact, from Eq.~(\ref{mbc}) and
(\ref{oba1}) we have
\begin{equation}
\frac{\tilde{M}^2_{Z_2}}{M^2_W}=\frac{(1-2s^2_W)(4+\overline{v}^2_{_W})
+4s^4_W(1-2\overline{v}^4_{_W})}{3c^2_W(1-4s^2_W)}\;
\frac{1}{\overline{v}^2_{_W}},
\label{goldratio}
\end{equation}
and we see that for $\overline{v}_{_W}=1$ (the 3-3-1 scale is equal to the
electroweak scale) we obtain $M_{Z_2}=3.77 M_W$. Of course for lower values of
$\overline{v}_{_W}$, $Z_2$ is heavier, for instance for $\overline{v}_{_W}=0.25$
we have $M_{Z_2}=18.36 M_W$. We recall that since $v_\chi$ does not contribute
to the $W$ mass it is not constrained by the 246 GeV upper bound. Thus,
independently if $\overline{v}^2_W$ is larger, smaller or equal to 1, we see
that the charged vector boson  $V$ is heavier than $U$. Using Eqs.~(\ref{mbc})
we see that in general $M^2_V-M^2_U=(g^2/4)(v^2_\eta-v^2_\rho)$ and, after using
Eqs.~(\ref{sol3}) and (\ref{con1}) we can write
\begin{equation}
\frac{\Delta M}{M_W}=\frac{(\tilde{M}_V^2-\tilde{M}_U^2)^{1/2}}{M_W}=
\left(3\tan^2\theta_W-\frac{2v^2_s}{v^2_W}\right)^{1/2}
\leq\sqrt{3}\tan\theta_W,
\label{delta}
\end{equation}
with $\Delta M/M_W\approx0.94$ for $v_s=0$, as is the case in the model with
heavy leptons~\cite{pt}. Notice that in this minimal model the scalar sextet
is introduced only to give the correct mass to the charged leptons, so it is not
necessarily a large VEV, $v_s\approx2$ GeV may be enough.

We see that, very impressing, the solution to the condition
$\rho_1=1$ given in Eq.~(\ref{sol3}), implies that the $Z$ has
\textit{exactly} the same mass than the respective vector boson of
the SM. Thus, unlike the case when $v_\chi\to\infty$ this is far
from being a trivial condition. It implies that the expressions for
neutral current parameters with $Z_1$ and $Z_2$, that usually have
been considered only approximately (valid when terms of the order
$v^2_W/v^2_\chi$ are neglected), are now exact expressions that
depend only on the weak mixing angle $\theta_W$. We will show below
that the condition in Eq.~(\ref{sol3}) is protected by an accidental
symmetry and when it is used the VEVs are not arbitrary anymore, up
to the sum in Eq.~(\ref{con1}).

Notice also that, from Eq.~(\ref{az1z2}), at a high energy $\mu$
when $s^2_W(\mu)=1/4$,  the photon $A_\mu$ and $Z_{1\mu}$ are the
gauge bosons of an $SU(2)_L\otimes U(1)_Y$ symmetry while
$Z_{2\mu}\equiv B_\mu$, where $B_\mu$ is the gauge boson of the
Abelian factor $U(1)_X$. It means that at high energies the product
$SU(2)\otimes U(1)\subset SU(3)$ decouples from $U(1)_X$ and this
may happen even at the electroweak scale.

\subsection{SM limit in the model with heavy leptons}
\label{subsec:smhl}

In this model all happens in the same way that in the previous one, except that
$v_s=0$ and then the condition in Eq.~(\ref{sol3}) is valid, and
which implies also, instead of (\ref{con1}),~$v^2_\eta~=~[(1~+~2s^2_W)/2c^2_W]v^2_W$.
By using $s^2_W=0.2312$~\cite{pdg} we obtain
that $v_\rho\approx54$ GeV and $v_\eta\approx240$ GeV. It means that in
this case there is only a free VEV: $v_\chi$. In particular, in this model
without sextet we have $\Delta M/M_W\approx0.94$ as shown in the previous
sub-section.

\subsection{SM limit in the model with right-handed neutrinos}
\label{subsec:smnr}

In this case we have the definition
\begin{equation}
\rho_1\equiv
\frac{c^2_WM^2_{Z_1}}{M^2_W}=\frac{2c^2_W}{\overline{v}_{_W}^2}
\,A(1-R), \label{rho2}
\end{equation}
with $A$ and $R$ given in Eqs.~(\ref{ap}) and (\ref{rp}),
respectively. Thus, in this model the condition $\rho_1=1$ implies
\begin{equation}
\tilde{\overline{v}}^2_\rho=\frac{(1-2s^2_W)}{2c^2_W}\,\overline{v}^2_{W},
\label{sol3nr}
\end{equation}
which gives the numerical values of $\tilde{v}_\rho=145.5$ GeV and
$\tilde{v}_\eta=198.4$ GeV. We have verified that when this relation
is used in the exact expressions for the mass eigenstates, in basis
$W^3_\mu,W^8_\mu,B_\mu$ given in Eq.~(\ref{w3w8bnd}), do not depend
on the VEVs structure, only on $s_W$, as in Eq.~(\ref{az1z2}).

\section{Vector and axial neutral current parameters}
\label{sec:nc}

Next, we will study the effect of the relations in Eqs.~(\ref{sol3})
and (\ref{sol3nr}) on the neutral current parameters of these
models. In this vain we will parameterize the neutral currents in
the 3-3-1 models considered above, as follows:
\begin{eqnarray}
{\cal{ L}}^{NC}_{331}=- \frac{g}{2c_W}\sum_i  \overline{ \psi_i}
\gamma^\mu \left[(g^i_V-g^i_A \gamma_5 ) Z_{1\mu} + (f^i_V-
f^i_A\gamma_5 )Z_{2\mu}\right]\psi_i,
\label{nc}
\end{eqnarray}
when the exact forms for $Z_1$ and $Z_2$, obtained by inverting
Eqs.~(\ref{w3w8b}) and (\ref{w3w8bnd}), are used in defining
$g^i_{V,A}$ and $f^i_{V,A}$. We recall that the fermions $\psi_i$
are all still symmetry eigenstates.

Below we will write the analytical exact expressions for the neutral
current parameters $g^{i}_{V,A}$ and $f^{i}_{V,A}$, showing
explicitly that they depend on the VEVs in a complicated way. But
when the conditions in Eqs.~(\ref{sol3}) and (\ref{sol3nr}) are used
in these expressions, in the respective model, we obtain for the
case of the known fermions $g^{i}_{V,A}\equiv g^{i(SM)}_{V,A}$, and
$f^{i}_{V,A}= f^i_{V,A}(s_W)$, i.~e., these parameters  depend only
on the electroweak mixing angle.

\subsection{Neutral current parameters in the minimal 3-3-1 model }
\label{subsec:ncmm}

The reduction of the complicated expression, depending on the VEVs
for the eigenstates, in Eq.~(\ref{w3w8b}) to those in
Eq.~(\ref{az1z2}) which depend only on $s_W$, is not a trivial
result. Moreover, we shall calculate the neutral current parameters
using the full expressions in (\ref{w3w8b}) and then use the
condition (\ref{sol3}). The result is that, independently of the
value of $v_\chi$, we obtain \textit{exactly} the parameters of the
SM $Z$-boson for those particles that are common with the 3-3-1
models, and that the extra non-standard particles have vector and
axial-vector neutral current parameters that do not depend on the VEVs,
but only on $\theta_W$. This implies that the effects of the extra
neutral currents due to $Z_2$ are only constrained by the $Z_2$ mass.
Moreover, in the first two models the $Z^\prime\equiv Z_2$ boson is
leptophobic~\cite{dumm97}.

First consider the leptonic sector. The coupling of the  neutrinos are:
\begin{eqnarray}
& & g^\nu_V=g^\nu_A=
-N_1\,\frac{c_W}{3}  \left(1-6 m^2_2 -  3\overline{v}_\rho^2
+4\overline{v}_{_W}^2\right),
\nonumber\\
& &f^\nu_V= f^\nu_A=
-N_2\,\frac{c_W}{3} \left(1-6 m^2_1 - 3 \overline{v}_\rho^2
+4\overline{v}^2_{_W}\right),
\label{gvgal1m1}
\end{eqnarray}
where $m^2_1$ and $m^2_2$ are defined in Eq.~(\ref{mz1z2mm}),
and $N_1$ and $N_2$ are defined in Eq.~(\ref{norm}).

For the case of the known charged leptons:
\begin{eqnarray}
& & g^l_V=  -N_1\, c_W \left(1-\overline{v}_\rho^2 \right),
\quad
g^l_A= N_1\,\frac{c_W}{3} \left(1-6m^2_2-3\overline{v}^2_W
\right),
\nonumber\\
& & f^l_V=  -N_2\,c_W  \left(1-\overline{v}_\rho^2
\right),
\quad
 f^l_A= N_2\,\frac{c_W}{3} \left(1- 6 m^2_1
-3\overline{v}_{_W} ^2+ 4\overline{v}_\rho^2    \right).
\label{gvgalexamm}
\end{eqnarray}

In the known quark sector we have the exact $g^q_{V,A}$ parameters
given by
\begin{eqnarray}
&&g^{u_m}_V=\frac{N_1}{h^2(s^2_W)}\,\frac{c_W}{3}\,[1+(3m^2_2-\overline{v}^2_{_W})\,
h^2(s^2_W) +2s^2_W(\overline{v}^2_\rho-3)],
\nonumber \\
&&g^{u_m}_A=\frac{N_1}{h^2(s^2_W)}\frac{c_W}{3}\,\,[1+(3m^2_2-2\overline{v}^2_{_W})\,
h^2(s^2_W)+ 2s^2_W(1+3\overline{v}^2_\rho)],
\nonumber \\
&&
g^{u_3}_V=-\frac{N_1}{h^2(s^2_W)}\,\frac{c_W}{3}\,[1-6m^2_2\,h^2(s^2_W)
-3\overline{v}^2_\rho
+4s^2_W(1+\overline{v}^2_\rho-4\overline{v}^2_{_W})],\nonumber \\
&&
g^{u_3}_A=
-N_1\,\frac{c_W}{3}\,(1-6m^2_2-3\overline{v}^2_\rho+4\overline{v}^2_{_W}),
\nonumber \\
&& g^{d_m}_V=
\frac{N_1}{h^2(s^2_W)}\,\frac{c_W}{3}\,[1+(4\overline{v}^2_{_W}-6m^2_2)\,h^2(s^2_W)-
(3-8s^2_W)\overline{v}^2_\rho],
\nonumber \\
&&
g^{d_m}_A=-g^{u_3}_A,\nonumber \\
&& g^{d_3}_V=-\frac{N_1}{h^2(s^2_W)}\,\frac{c_W}{3}\, [1+(
4\overline{v}^2_{_W}+3m^2_2)\,h^2(s^2_W)-2s^2_W(1+\overline{v}^2_\rho)],
\nonumber \\
&& g^{d_3}_A=
-\frac{N_1}{h^2(s^2_W)}\,\frac{c_W}{3}\,[1+(3m^2_2-2\overline{v}^2_{_W})\,
h^2(s^2_W)+2s^2_W(1-3\overline{v}^2_\rho)]. \label{kqmmexa}
\end{eqnarray}

Finally, in the exotic quarks sector we have that the exact $g^j_{V,A}$ parameters
are
\begin{eqnarray}
&&g^{j_m}_V=
-\frac{N_1}{h^2(s^2_W)}\,\frac{c_W}{3}\,[2+(2\overline{v}^2_{_W}-3m^2_2)\,h^2(s^2_W)-
3\overline{v}^2_\rho-2s^2_W(9-11\overline{v}^2_\rho)],\nonumber \\
&& g^{j_m}_A=-\frac{N_1}{h^2(s^2_W)}\,\frac{c_W}{3}\,[ 2+(2
\overline{v}^2_{_W}-
3m^2_2)\,h^2(s^2_W)-3\overline{v}^2_\rho-2s^2_W(1-3\overline{v}^2_\rho)], \nonumber \\
&& g^J_V=
\frac{N_1}{h^2(s^2_W)}\,\frac{c_W}{3}\,[2+(2\overline{v}^2_{_W}
-3m^2_2)\,h^2(s^2_W)- 3\overline{v}^2_\rho
-2s^2_W(11-13\overline{v}^2_\rho)],\nonumber \\
&&
g^J_A=\frac{N_1}{h^2(s^2_W)}\,\frac{c_W}{3}\,[2+(2\overline{v}^2_{_W}
-3m^2_2)\,h^2(s^2_W) -3\overline{v}^2_\rho
-2s^2_W(1-3\overline{v}^2_\rho)], \label{gvgaeqexamm}
\end{eqnarray}
where we have denoted $h(s^2_W)=+[1-4s^2_W]^{1/2}$. The parameters
of quarks to $Z_{2\mu}$, $f^q_{V,A}$, are obtained from those in
Eqs.~(\ref{kqmmexa}) and (\ref{gvgaeqexamm}) by replacing $N_1\to
N_2$ and $m^2_2\to m^2_1$ and we will not write them explicitly.

When the relation in Eq.~(\ref{sol3}) is used in the $g^i_{V,A}$ and
$f^i_{V,A}$ parameters  in Eqs.~(\ref{gvgalexamm}) -
(\ref{gvgaeqexamm}), we obtain (here we omitted the tilde
($\tilde{\phantom{x}}$) in all the expressions on the left side)
\begin{eqnarray}
 & & g^\nu_V =g^\nu_A =  \frac{1}{2},  \qquad\qquad\qquad
f^\nu_V = f^\nu_A = -\frac{\sqrt3}{6}\,h(s^2_W),
\nonumber \\
& & g^l_V  =  -\frac{1}{2}+2s^2_W,  \quad\qquad\qquad g^l_A  =
-\frac{1}{2}, \nonumber \\ & & f^l_V = -f^l_A= -
\frac{\sqrt3}{6}\,h(s^2_W), \label{gvgal}
\end{eqnarray}
in the lepton sector, while in the known quark sector we have:
\begin{eqnarray}
 & & g^{u}_V =\frac{1}{2}-\frac{4}{3}\,s^2_W \quad g^{u}_A =\frac{1}{2},
 \quad u=u_1,u_2,u_3;\nonumber \\
 &&
f^{u_m}_V = \frac{1}{2\sqrt{3}}\,\frac{1-6s^2_W}{h(s^2_W)}, \quad
f^{u_m}_A =\frac{1}{2\sqrt{3}}\,\frac{1+2s^2_W}{h(s^2_W)},\quad
m=1,2;
\nonumber \\
&& f^{u_3}_V=-\frac{1}{2\sqrt{3}}\,\frac{1+4s^2_W}{h(s^2_W)},\quad
f^{u_3}_A=-\frac{1}{\sqrt3}\,h(s^2_W),
\nonumber \\
&&
g^{d}_V  =  -\frac{1}{2}+\frac{2}{3}s^2_W,\quad
g^{d}_A  =  -\frac{1}{2}, \quad d=d_1,d_2,d_3;
\nonumber \\
&& f^{d_m}_V=\frac{1}{2\sqrt{3}\,h(s^2_W)},\quad
f^{d_m}_A=\frac{h(s^2_W)}{2\sqrt3},\,\quad m=1,2; \nonumber \\
&& f^{d_3}_V=-\frac{1}{2\sqrt3}\,\frac{1-2s^2_W}{h(s^2_W)} , \quad
f^{d_3}_A = -\frac{1}{2\sqrt3}\,\frac{1+2s^2_W}{h(s^2_W)},
\label{gvgaud}
\end{eqnarray}
and, finally in the exotic quark sector:
\begin{eqnarray}
& & g^{j_m}_V = \frac{8}{3}\,s^2_W,  \quad g^{j_m}_A = 0,\quad
 f^{j_m}_V =-\frac{1}{\sqrt3}\,\frac{1-9s^2_W}{h(s^2_W)},\quad
f^{j_m}_A =-\frac{1}{\sqrt3}\,\frac{c^2_W}{h(s^2_W)}, \quad m=1,2;
\nonumber \\
& & g^J_V = -\frac{10}{3}\,s^2_W,  \quad g^J_A = 0,\quad
 f^J_V =\frac{1}{\sqrt3}\,\frac{1-11s^2_W}{h(s^2_W)},\quad
f^J_A =\frac{1}{\sqrt3}\,\frac{c^2_W}{h(s^2_W)}. \label{gvgaeq}
\end{eqnarray}
Notice that, since all fields in Eq.~(\ref{nc}) are symmetry
eigenstates, from the parameters in Eqs.~(\ref{gvgal}) to
(\ref{gvgaeq}) we see that in the leptonic sector there is not FCNC
neither with $Z_{1\mu}$ nor with $Z_{2\mu}$ and in the quark sector
there are FCNC only coupled to $Z_{2\mu}$~\cite{dpp}. Notice also
that the exotic quarks have pure vectorial couplings with
$Z_{1\mu}$.

\subsection{Neutral current parameters in the model with heavy leptons}
\label{subsec:nchl}

Let us consider the parameters in the neutral currents coupled with
$Z_{1\mu}$ and $Z_{2\mu}$. For the neutrinos they are:
\begin{eqnarray}
& & g^\nu_V=g^\nu_A=
N_1\,c_W \left( 2 m^2_2 +  \overline{v}_\rho^2
-\frac{4}{3}\overline{v}_{_W} ^2- \frac{1}{3}   \right),
\nonumber\\
& &f^\nu_V= f^\nu_A=
N_2\,c_W \left( 2 m^2_1 +  \overline{v}_\rho^2
-\frac{4}{3}\overline{v}_{_W}^2- \frac{1}{3}   \right).
\label{gvgan}
\end{eqnarray}

For the case of the known charged leptons:
\begin{eqnarray}
& & g^l_V=  -N_1\,c_W \left[  m^2_2
+\frac{1}{3}(1-2\overline{v}_{_W}^2)- \frac{2s^2_W}{h^2(s^2_W)}
(1-\overline{v}_\rho^2 ) \right],
\nonumber\\
& & g^l_A= -N_1\,c_W \left[  m^2_2 +\frac{1}{3}(1-2\overline{v}_{_W}
^2)+ \frac{2s^2_W}{h^2(s^2_W)} (1-\overline{v}_\rho^2 ) \right],
\nonumber\\
& & f^l_V=  -N_2\,c_W  \left[  m^2_1
+\frac{1}{3}(1-2\overline{v}_{_W} ^2)- \frac{2s^2_W}{h^2(s^2_W)}
(1-\overline{v}_\rho^2 )   \right],
\nonumber\\
& & f^l_A= -N_2\,c_W \left[  m^2_1 +\frac{1}{3}(1-2\overline{v}_{_W}
^2)+ \frac{2s^2_W }{h^2(s^2_W)} (1-\overline{v}_\rho^2 )   \right].
\label{gvgal2}
\end{eqnarray}

For the exotic heavy charged leptons we have:
\begin{eqnarray}
& & g^E_V=  -N_1\,c_W \left[  m^2_2
+\overline{v}_\rho^2-\frac{2}{3}(1+\overline{v}_{_W} ^2)+
\frac{2s^2_W }{h^2(s^2_W) }\, (1-\overline{v}_\rho^2 )   \right],
\nonumber\\
& & g^E_A= -N_1\,c_W \left[  m^2_2
+\overline{v}_\rho^2-\frac{2}{3}(1+\overline{v}_{_W} ^2)-
\frac{2s^2_W}{h^2(s^2_W)}\, (1-\overline{v}_\rho^2 )   \right],
\nonumber\\
& & f^E_V=  -N_2\,c_W \left[  m^2_1
+\overline{v}_\rho^2-\frac{2}{3}(1+\overline{v}_{_W} ^2)+
\frac{2s^2_W }{h^2(s^2_W) } \,(1-\overline{v}_\rho^2 )   \right],
\nonumber\\
& & f^E_A= - N_2\,c_W \left[  m^2_1
+\overline{v}_\rho^2-\frac{2}{3}(1+\overline{v}_{_W}^2)-
\frac{2s^2_W }{h^2(s^2_W) }\, (1-\overline{v}_\rho^2 )   \right].
\label{gvgaE}
\end{eqnarray}
Finally, in the quark sector the parameters are the same as in
Eqs.~(\ref{kqmmexa}) and (\ref{gvgaeqexamm}) or, after using
(\ref{sol3}), in (\ref{gvgaud}) and (\ref{gvgaeq}).

If we substitute the condition in Eq.~(\ref{sol3}) in all the
$g^i_{V,A}$ and $f^i_{V,A}$ parameters, given in Eqs.~(\ref{gvgan})
-- (\ref{gvgaE}), we obtain (here also we are omitting the tilde in
all the expressions on the left side):

\begin{eqnarray}
 & & g^\nu_V =g^\nu_A =  \frac{1}{2},  \hspace{2.5 cm}
f^\nu_V = f^\nu_A = -\frac{\sqrt3}{6}\, h(s^2_W),
 \nonumber\\
& & g^l_V  =  -\frac{1}{2}+2s^2_W,  \qquad\qquad
g^l_A  =  -\frac{1}{2},
 \nonumber\\ & & f^l_V =
- \frac{\sqrt3}{6} \frac{(1-10s^2_W) }{h(s^2_W)}, \quad f^l_A = -
\frac{\sqrt3}{6} \frac{(1+2s^2_W) }{h(s^2_W)},
 \nonumber\\
& & g^E_V = -2s^2_W,  \quad\quad g^E_A = 0,
 \nonumber\\
& & f^E_V = \frac{\sqrt3}{3} \frac{(1-7s^2_W) }{h(s^2_W)}, \quad
f^E_A =
 \frac{\sqrt3}{3} \frac{c^2_W }{h(s^2_W)},
\label{gvgaEapzl}
\end{eqnarray}
and we see that for neutrinos and the usual known charged leptons
the neutral current parameters in the $Z_{1\mu}$ interactions are
exactly the same as those in the SM at the tree level. Notice also
that only the neutrinos have leptophobic interactions with
$Z_{2\mu}$ in this model.

\subsection{Neutral current parameters in the model with right-handed
neutrinos}
\label{subsec:ncnr}

In this model we have also obtained the exact neutral current
parameters in both sectors $Z_{1\mu}$ and $Z_{2\mu}$, which as in
the previous models are denoted by $g^i_{V,A}$ and $f^i_{V,A}$,
respectively. For the neutrinos we obtain:
\begin{eqnarray}
& & g^\nu_V=N_1\,c_W [ 6 m^2_2 -3( \overline{v}^2_{_W}
-\overline{v}^2_\rho)-3],
\nonumber\\
& & g^\nu_A= \frac{N_1c_W}{k^2(s^2_W)}\,[1+\overline{v}_\rho^2 -
\overline{v}^2_{_W}],
\nonumber\\
& &f^\nu_V=
 N_2\,c_W [ 6 m^2_1 -3( \overline{v}_{_W} ^2
-\overline{v}_\rho^2)-3],
\nonumber\\
& &  f^\nu_A=  \frac{N_2c_W}{k^2(s^2_W)}\,[ 1+\overline{v}^2_\rho -
\overline{v}^2_{_W}], \label{gvgalrhn1}
\end{eqnarray}
where we have denoted $k(s^2_W)=+[1-(4/3)s^2_W]^{1/2}$. And, in the
charged lepton sector:
\begin{eqnarray}
& & g^l_V=  -\frac{N_1c_W}{k^2(s^2_W)}\, [ 1+\overline{v}_\rho^2-
\overline{v}^2_{_W}]h^2(s^2_W),
\nonumber\\
& & g^l_A= - \frac{N_1c_W}{k^2(s^2_W)}\,[ 1+\overline{v}^2_\rho-
\overline{v}^2_{_W}],
\nonumber\\
& & f^l_V=  -\frac{N_2c_W}{k^2(s^2_W)}\,[ 1+\overline{v}^2_\rho-
\overline{v}^2_{_W}]h^2(s^2_W),
\nonumber\\
& & f^l_A= -\frac{N_2c_W}{k^2(s^2_W)}\,[ 1+\overline{v}_\rho^2-
\overline{v}_{_W} ^2]. \label{gvgalrhn2}
\end{eqnarray}

Using the condition (\ref{sol3nr}) in Eqs.~(\ref{gvgalrhn1}) and
(\ref{gvgalrhn2}) we obtain (here again we have omitted the tilde
($\tilde{\phantom{x}}$) in all the expressions on the left side)
\begin{eqnarray}
& & g^\nu_V =g^\nu_A =   \frac{1}{2},
\nonumber \\
 & & f^\nu_V =  -\frac{\sqrt3}{2}\,k(s^2_W), \quad
 f^\nu_A =
\frac{ \sqrt3}{ 6}\,\frac{ 1}{ k(s^2_W)},
\nonumber\\
& & g^l_V  =  -\frac{1}{2}+2s^2_W,  \quad
g^l_A  =  -\frac{1}{2},
\nonumber\\
& & f^l_V=\frac{\sqrt3}{6}\, \frac{h^2(s^2_W)}{k(s^2_W)}, \quad
f^l_A = -\frac{\sqrt3}{6}\, \frac{1}{ k(s^2_W)}, \label{gvgalrhn3}
\end{eqnarray}
and we see that, again, the parameters in the $Z_{1\mu}$ currents
coincide in an exactly way with those of the $Z$ in the SM. Notice
also that in this model $Z_{2\mu}$ is leptophobic only with charged
leptons. The reduction of the exact parameters of the known fermions
with $Z_{1\mu}$ to those of the SM, when the condition
(\ref{sol3nr}) is used, also occurs in the quark sector but here we
will not write them explicitly.

\section{Custodial symmetry}
\label{sec:custodial}

We can rewrite Eqs.~(\ref{sol3}) which is valid for the models of
Secs.~\ref{subsec:mm} and \ref{subsec:hl}, as
\begin{equation}
g\;\frac{ \tilde{v}_\rho}{\sqrt2}= \frac{\sqrt{1-4s^2_W}}{c_W}\;M_W,
\label{sol3b}
\end{equation}
and Eq.~(\ref{sol3nr}) which is valid for the model of
Sec.~\ref{subsec:rn}, as
\begin{equation}\label{sol3nrb}
g\frac{\tilde{v}_\rho}{\sqrt2}=\frac{\sqrt{1-2s^2_W}}{c_W}\;M_W.
\end{equation}
These are like the Goldberger-Treiman relation~\cite{gt58} in the
sense that their validity imply, as we will show below, an
approximate global symmetry of the models and all quantities
appearing in these relations can be measured independently: the $W$
mass $M_W$, the sine of the weak mixing angle $\sin\theta_W$, the
$SU(3)_L$ coupling constant, $g$, and the VEV of one of the
triplets, say, $v_\rho$. In fact, all but $\tilde{v}_\rho$, are
already well known. However, cross sections of several processes,
for instance $e^+e^-\to ZH$ where $H$ is a neutral Higgs scalar
transforming as doublet of $SU(2)$, are sensitive to the value of
$v_\eta$ (or $v_\rho$)~\cite{cieza}. So, in principle it is possible
to verify if Eq.~(\ref{sol3b}), or Eq.~(\ref{sol3nrb}), is satisfied
and if the 3-3-1 symmetry can be implemented near the electroweak
scale.

We can study the ``stability" of the full expressions for $\rho_1$
in Eq.~(\ref{rho1}) using the full expression for $M_{Z_1}$ given in
Eqs.~(\ref{mz1z2mm}), with Eqs.~(\ref{amm}) and (\ref{rmm}) for the
case of the minimal and heavy lepton models, and Eqs.~(\ref{ap}) and
(\ref{rp}) for the case of the minimal model and for the model with
right-handed neutrinos. We have analyzed how the condition
$\rho_1=1$ varies when we change arbitrarily $v_\rho$. We expand the
value of $v_\rho$ as
$\overline{v}^\prime_\rho=(1+x)\tilde{\overline{v}}_\rho$ and
substituting $\overline{v}^\prime_\rho$ in Eq.~(\ref{rho1}) and
expanding in $x\lesssim1$ we obtain
\begin{eqnarray}
\rho_1\approx 1-0.0025x^2+0.00012x^3+\cdots,\;\;\textrm{for}\;\;  \overline{v}_W
=1;\\\rho_1\approx 1-0.00024x^2+1.8\times10^{-6}x^3+\cdots,\;\; \textrm{for}\;\;
\overline{v}_W=0.1,
\label{xismm}
\end{eqnarray}
for the minimal model, and
\begin{eqnarray}
\rho_1\approx 1-0.3497x^2+0.1051x^3+\cdots,\;\;\textrm{for}\;\;  \overline{v}_W=1;\\
\rho_1\approx 1-0.0131x^2+0.0001x^3+\cdots,\;\; \textrm{for}\;\;\overline{v}_W=0.1,
\label{xisnr}
\end{eqnarray}
for the model with right-handed neutrinos. Thus, we see that the
minimal model is more stable, in the sense discussed above, than the
model with right-handed neutrinos with respect to the departure of
$v_\rho$ from $\tilde{v}_\rho$, i. e., from the condition
(\ref{sol3b}) or (\ref{sol3nrb}), respectively. For instance,
$\overline{v}_W=1$, for the minimal model we have that a 20\%
($x=0.2$) depart from the condition (\ref{sol3b}) the value of
$\rho_1$ is only affected by 0.01 \%, while for the model with
right-handed neutrinos and (\ref{sol3nrb}), for the same value of
$x$, the respective $\rho_1$ changes 1.28 \%. This suggests that,
when both 3-3-1 models were embedded in a $SU(4)_L\otimes U(1)_N$
model~\cite{su4} which has three real neutral vector bosons, the
$SU(3)$ subgroup which contains the SM's $Z$ should be the minimal
3-3-1 model considered in Sec.~\ref{subsec:mm}.

In the SM the fact that $\rho_0=1$ is a consequence of an
approximate (accidental)  $SU(2)_{L+R}$ global symmetry named
``custodial symmetry"~\cite{custodial,will}. This custodial symmetry
is exact when $g^\prime=0$ ($\sin^2\theta_W=0$) which implies
$M_W=M_Z$ since in this limit $W^+,W^-,Z$ form a triplet of this
unbroken global symmetry. Also, due to the unbroken $SU(2)_{L+R}$ in
the $g^\prime\to0$ limit, radiative corrections to the $\rho_0$
parameter due to gauge and Higgs bosons must be proportional to
$g^{\prime 2}$~\cite{will}. We try to understand this situation in
the context of 3-3-1 models, by showing that these models have also
an approximate global $SU(2)_{L+R}$.

Let us consider the model of Sec.~\ref{subsec:hl}. We can decompose
the triplets as ${\bf3}={\bf2}+{\bf1}$ under $SU(2)_L\otimes
U(1)_Y$. In particular, the scalar triplets we have used can be
written as
\begin{equation}
\varphi=H_\varphi+s_\varphi,
\label{dubletos}
\end{equation}
where $\varphi=\eta,\rho,\chi$. Under $SU(2)_L\otimes U(1)_Y$,
$H_\eta,H_\rho,H_\chi$ transform as $ ({\bf2},-1/2), ({\bf2},1/2),
({\bf2},-3)$, respectively, and $s_\eta,s_\rho,s_\chi$ as
$({\bf1},+2), ({\bf1},+4), ({\bf1},0)$, respectively. Next, we
define four 2-doublet
\begin{equation}
\Phi_{\zeta\zeta}=\frac{1}{\sqrt2}(\tilde{H}_\zeta\,H_\zeta),\quad\quad
\Phi_{\rho\eta}=\frac{1}{\sqrt2}(H_\rho\,H_\eta), \label{2D}
\end{equation}
where  $\zeta=\eta,\rho,\chi$, and $\tilde{H}=\epsilon H^*$ . We can
write the more general scalar potential invariant under 3-3-1 as
\begin{equation}
V(\eta,\rho,\chi)=V(\Phi_{\varphi\varphi^\prime},s_\varphi)+f\eta\rho\chi+H.c.,
\label{truque}
\end{equation}
where
\begin{eqnarray}
V(\Phi_{\varphi\varphi^\prime},s_\varphi)&=
&\mu^2_\eta[\textrm{Tr}(\Phi^\dagger_{\eta\eta}\Phi_{\eta\eta})
+s^\dagger_\eta s_\eta]+
\mu^2_\rho[\textrm{Tr}(\Phi^\dagger_{\rho\rho}\Phi_{\rho\rho})+s^\dagger_\rho s_\rho]+
\mu^2_\chi [\textrm{Tr}(\Phi^\dagger_{\chi\chi}\Phi_{\chi\chi})+s^\dagger_\chi s_\chi]
\nonumber \\ &+&
\lambda_1[\textrm{Tr}(\Phi^\dagger_{\eta\eta} \Phi_{\eta\eta})+s^\dagger_\eta s_\eta]^2
+\lambda_2[\textrm{Tr}(\Phi^\dagger_{\rho\rho} \Phi_{\rho\rho})+s^\dagger_\rho s_\rho]^2 +
\lambda_3[\textrm{Tr}(\Phi^\dagger_{\chi\chi} \Phi_{\chi\chi})+s^\dagger_\chi s_\chi]^2
\nonumber \\ &+&
\lambda_4[\textrm{Tr}(\Phi^\dagger_{\chi\chi} \Phi_{\chi\chi})+s^\dagger_\chi s_\chi]
[\textrm{Tr}(\Phi^\dagger_{\eta\eta} \Phi_{\eta\eta})+s^\dagger_\eta s_\eta]
 +\lambda_5 [\textrm{Tr}(\Phi^\dagger_{\chi\chi} \Phi_{\chi\chi})+s^\dagger_\chi s_\chi]
[\textrm{Tr}(\Phi^\dagger_{\rho\rho} \Phi_{\rho\rho})\nonumber \\ &+&s^\dagger_\rho s_\rho]
+
\lambda_6 [\textrm{Tr}(\Phi^\dagger_{\rho\rho} \Phi_{\rho\rho})+s^\dagger_\rho s_\rho]
[\textrm{Tr}(\Phi^\dagger_{\eta\eta} \Phi_{\eta\eta})+s^\dagger_\eta s_\eta]+
\{\lambda_7 [\textrm{Tr}(\Phi^\dagger_{\chi\chi} \Phi_{\eta\eta})+s^\dagger_\chi s_\eta]
\nonumber \\ &\cdot&
[\textrm{Tr}(\Phi^\dagger_{\eta\eta} \Phi_{\chi\chi})+s^\dagger_\eta s_\chi]+\lambda_8
[\textrm{Tr}(\Phi^\dagger_{\chi\chi} \Phi_{\rho\rho})+s^\dagger_\chi s_\rho]
[\textrm{Tr}(\Phi^\dagger_{\rho\rho} \Phi_{\chi\chi})+s^\dagger_\rho s_\chi]
\nonumber \\ &+&\lambda_9 [\textrm{Tr}(\Phi^\dagger_{\rho\rho} \Phi_{\eta\eta})
+s^\dagger_\rho s_\eta]
[\textrm{Tr}(\Phi^\dagger_{\eta\eta} \Phi_{\rho\rho})+s^\dagger_\eta s_\rho]
\nonumber \\ &+&
\lambda_{10}[\textrm{Tr}(\Phi^\dagger_{\chi\chi} \Phi_{\eta\eta})
+s^\dagger_\chi s_\eta]
[\textrm{Tr}(\Phi^\dagger_{\rho\rho} \Phi_{\eta\eta})+s^\dagger_\rho s_\eta] +H.c.\}.
\label{potencialsu2}
\end{eqnarray}

The full scalar potential in Eq.~(\ref{truque}) is invariant under
the 3-3-1 symmetry, but the part
$V(\Phi_{\varphi\varphi^\prime},s_\varphi)$ in
Eq.~(\ref{potencialsu2}) is also invariant under
$\Phi_{\varphi\varphi^\prime}\to L\Phi_{\varphi\varphi^\prime}$,
$\Phi_{\varphi\varphi^\prime}\to \Phi_{\varphi\varphi^\prime}
R^\dagger$, and $s_\varphi\to s_\varphi$ i. e., $SU(2)_L\otimes
SU(2)_R$ as in the standard electroweak model, when $\sin\theta_W=0$
($g^\prime=0$ in $SU(2)_L\otimes U(1)_Y$ models). Notwithstanding,
the trilinear term in Eq.~(\ref{truque}), $f(\eta\rho\chi)$, breaks
softly this global custodial symmetry. It means that this symmetry
is realized only in the limit $f\to0$. When the Higgs fields acquire
vacuum expectation values we have
\begin{equation}
\langle\Phi_{\eta\rho}\rangle=\frac{1}{\sqrt2}\,\left( \begin{array}{cc}
v_\eta & 0 \\
0 &v_\rho \\
\end{array}\right),
\label{c6nova}
\end{equation}
breaking $SU(2)_L\otimes SU(2)_R\to SU(2)_{L+R}$, if $v_\eta=v_\rho$
in Eq~(\ref{potencialsu2}). Notice also
that $SU(3)_L$ transformations mix the components of the bi-doublets with the singlets,
thus breaking also the custodial symmetry. Moreover, the covariant derivative implies
the mixing between $Z$ and $Z^\prime$ which breaks explicitly the custodial symmetry
unless $\sin\phi=0$.

This 3-3-1 model has an approximate $SU(2)_{L+R}$ symmetry as in the
Standard Model. This in fact does happen when Eq.~(\ref{sol3b})
(or~$\sin\phi=0$) holds and also $g_X=0$ (or $\sin\theta_W=0)$ then
$v_\eta=v_\rho=v_{_W}/\sqrt2$ and we really have the global
$SU(2)_{L+R}$ symmetry. This explains why $\sin\phi$ should be small
and may be generated by radiactive corrections only. In general, the
oblique parameters $T$ and $S$ constraint the mixing angle between
$Z$ and $Z^\prime$~\cite{pdg} with $\rho_0-1\simeq \alpha T$. The
$T$ parameter in 3-3-1 models has been calculated in
Ref.~\cite{inami}. Using their expressions but without the mixing
between $Z$ and $Z^\prime$ at the tree level (i. e., $\phi=0$ in
Eq.~(4.1) of \cite{inami}) we obtain, for example, $T=-0.1225$ for
$\overline{v}_{_W}=1$, and $T=-0.012$ for $\overline{v}_{_W}=0.25$.
We have also verified that $T\to 0$ when $v_\chi\to\infty$ and all
the values for $T$ with $\overline{v}_{_W}\leq1$ are within the
allowed interval~\cite{pdg}. This implies that the solution in
Eq.~(\ref{sol3b}) is not too much disturbed by radiative
corrections.

As a consistent verification, we note that if we substitute
Eq.~(\ref{sol3}) in the mass matrix of Eq.~(\ref{mnt})
we obtain
\begin{equation}
\hat{{\cal M}}^2_{(b=\sqrt{3})}=\frac{g^2}{4}v_\chi^2
\left(\begin{array}{ccc} \overline{v}^2_{_W} &
\sqrt{3}\,t^2_W\,\overline{v}^2_{_W}& -\frac{t_W\,h(s^2_W)\,
\overline{v}^2_{_W}}{c_W}
\\
\sqrt{3}\,t^2_W\,\overline{v}^2_{_W} & \frac{1}{3}\,(4+
\overline{v}^2_{_W}) & \frac{t_W [4c^2_W+h^2(s^2_W)
\,\overline{v}^2_{_W}]}{\sqrt{3}c_W\,h(s^2_W)}
\\
-\frac{t_W\,h(s^2_W)\,\overline{v}^2_{_W} }{c_W}
&\frac{t_W[4c^2_W+h^2(s^2_W)\,
\overline{v}^2_{_W}]}{\sqrt{3}c_W\,h(s^2_W)}
 &
\frac{2t^2_W}{h^2(s^2_W)} \,[2c^2_W+ h^2(s^2_W) \overline{v}^2_{_W}]
\end{array}
\right),
\label{masssol}
\end{equation}
where $t_W$ and $h(s^2_W)$ have already been defined in Sec.~\ref{subsec:smmm}
and \ref{subsec:ncmm}, respectively. The states $(W_{3\mu},W_{8\mu},B_\mu)$ are given in
terms of the mass eigenstates $(A_\mu,Z_\mu,Z^\prime_\mu)$, omitting the tilde in
the latter fields, as follows
\begin{eqnarray}
&&\tilde{W}^3_\mu=(s_W, -c_W,0),\quad
\tilde{W}^8_\mu=\left(-\sqrt{3}s_W,
-\sqrt{3}t_Ws_W, \frac{h(s^2_W)} {c_W}\right),\nonumber \\
&& \tilde{B}_\mu=\left(h(s^2_W),\,t_W h(4s^2_W),\sqrt{3}t_W\right),
\label{ufa}
\end{eqnarray}
which coincide with the inverse of Eq.~(\ref{az1z2}). Notice that
$\tilde{B}_\mu$ is almost $Z^\prime_\mu$, and when $s^2_W=1/4$ then
$\tilde{B}_\mu=Z^\prime_\mu$. The expressions in Eq.~(\ref{ufa}) is
consistent with those in Eq.~(\ref{az1z2}) after using the equation
Eq.~(\ref{sol3}).

In the limit $\sin\theta_W=0$  the mass square matrix in Eq.~(\ref{masssol})
reduces to
\begin{equation}
\tilde{{\cal M}}^2_{(b=\sqrt{3})}=\frac{g^2}{4}v_\chi^2
\left(\begin{array}{ccc}
\overline{v}^2_{_W} & 0& 0\\
0& \frac{1}{3}(4+\overline{v}^2_{_W}) & 0\\0&0&0
\end{array}
\right),
\label{ver}
\end{equation}
and we see that, using the relation (\ref{sol3b}) in
Eq.~(\ref{mbc}), in this limit $M_W=M_{Z_1}\equiv M_1$;
$M_U=M_V\equiv M_2$, where we have defined $M^2_1=g^2v^2_W/4$ and
$M^2_2=g^2(v^2_\chi+v^2_W/2)/4$. The photon of course continues
massless. These mass relations, valid in the limit $\sin\phi=0$ and
$\sin\theta_W=0$, are consequences of the custodial $SU(2)_{L+R}$
discussed above. Notice that in this limit, directly from
(\ref{ver}), or from (\ref{oba1}) with $s^2_W=0$, we have
$M^2_{Z^\prime}=g^2(4v^2_\chi+v^2_W)/12>M^2_2$. The custodial
symmetry appears also in the Yukawa sector as is shown in the
Appendix.

These models have also an approximate global $SU(3)_{L+R}$ symmetry
which, although badly broken, it may be useful for obtaining an
approximate but realistic mass spectra in the scalar sector as can
be seen in the Appendix~\ref{sec:a1}.

\section{Conclusions and discussions}
\label{sec:con}

Once the $v_\chi$ scale is arbitrary when Eq.~(\ref{sol3}) or (\ref{sol3nr})
are satisfied,
we can ask ourselves what about the experimental limit upon the extra
particles that appear in  the models. After all they
depend mainly on $v_\chi$, the scale at which the $SU(3)_L$ symmetry is
supposed to be valid. Here we will be concerned only with the minimal model
of Sec.~\ref{subsec:mm}.
Firstly, let us consider
the $Z^\prime$ vector boson. It contributes to the $\Delta M_K$ at the tree
level~\cite{dpp,liu94}. If this would be the only contribution to this parameter,
its experimental measurements constraint the quantity
\begin{equation}
({\cal O}^d_L)_{3d}({\cal O}^d_L)_{3s}\,\frac{M_Z}{M_{Z^\prime}},
\label{deltak}
\end{equation}
which must be of the order of $10^{-4}$ to have compatibility with
the measured $\Delta M_K$. This can be achieved with
$M_{Z^\prime}\sim 4$ TeV if we assume that the mixing matrix have a
Fritzsch-structure ${\cal O}^d_{Lij}=\sqrt{m_j/m_i}$~\cite{sher} or,
it is possible that the product of the mixing angles saturates the
value $10^{-4}$~\cite{dpp,liu94}, in this case $Z^\prime$ it is not
too much constrained and may have a mass near the electroweak scale.
More important is the fact that there is also in this model FCNC
mediated by neutral Higgs scalars which imply new contributions to
$\Delta M_K$ proportional to
\begin{equation}
({\cal O}^d_L)_{d3}\Gamma^d_{3\beta}({\cal O}_R)_{\beta s}\,\frac{M_Z}{M_H},
\label{deltak2}
\end{equation}
that involves the mass of the scalar $M_H$, the unknown ${\cal
O}^d_R$ matrix elements and also the Yukawa coupling $\Gamma^d$, so
that their contributions to $\Delta M_K$ may have opposite sing
relative to that of the $Z^\prime$ contribution. Thus, a  realistic
calculation of the $\Delta M_K$ in the context of 3-3-1 models has
to take into account these extra scalar contributions as well.
Hence, there is not strong constraints on the value the $Z^\prime$
mass in context of 3-3-1 models.

The model has also a doubly charged vector and four doubly charged
scalars. Muonium-antimuonium transitions would imply a lower bound
of 850 GeV on the mass of the doubly charged gauge bilepton,
$U^{--}_\mu$~\cite{willmann99}. However this bound depends on
assumptions on the mixing matrix in the lepton charged currents
coupled to $U^{--}_\mu$, and also it does not take into account that
there are in the model doubly charged scalar bileptons which also
contribute to that transition~\cite{pleitez00}. Concerning these
doubly charged scalars, model independent lower limits for their
masses are of the order of 100 GeV~\cite{d0}. From fermion pair
production at LEP and lepton flavor violating effects suggest a
lower bound of 750 GeV for the $U^{--}_\mu$ mass, but again it
depends on assumptions on the mixing matrix~\cite{tully}. Other
phenomenological analysis in $e^+e^-,e\gamma$ and $\gamma\gamma$
colliders assume bileptons with masses between 500 GeV and 1
TeV~\cite{dion,dion1,dion2}. The fine structure of muonium only
implies $M_U/g>215$ GeV~\cite{fujii} but also ignores the
contributions of the doubly charged scalars. Concerning the exotic
quark masses there is no lower limit for them but if they are in the
range of 200-600 GeV they may be discovered at the LHC~\cite{yara}.
Similarly, most of the searches for extra neutral gauge bosons are
based on models that do not have the same neutral current couplings
as those of the 3-3-1 models~\cite{babu}. Anyway we have seen that
even if $v_\chi=v_W$ the $Z^\prime_\mu$ has a mass of the order of
303 GeV. Of course, a value for $v_\chi$ of the order of 1 TeV could
be safer.

In view of this, we may conclude that there are not yet definitive
bounds on the masses of the extra degrees of freedom of the 3-3-1
models. Moreover, the $Z^\prime_\mu$ of the minimal 3-3-1 model has
interesting features that distinguishes this model from others
having also this sort of vector boson, as models with extra
dimensions and Little Higgs models. However because of its
leptophobic character it is not clear if it could be discovered at
the International Linear Collider~\cite{go}, probably the LHC may be
more efficient for searching it.

Finally, some remarks concerning 3-3-1 models in general. 1) The
existence of leptophobic neutral vector bosons were proposed in the
past to solve what would be anomalies in the weak precision data at
LEP, as $R_{b,c}$, see for example~\cite{docapeta}. Unlike other
sorte of models, the leptophobic boson is a prediction of the 3-3-1
models which already have interesting features. 2) The scalar sector
of the latter models has not deserved much attention in literature
and we think that phenomenological analysis as that in
Ref.~\cite{snowmass} should take into account these sort of models.
3) Usually in literature two models are mainly considered. The so
called ``minimal 3-3-1 model" in which the already known leptons
$(\nu_l\; l^-\, l^+)^T_L$ transform as $(\textbf{3},0)$ under
$SU(3)_L\otimes U(1)_X$~\cite{331}, and also the  ``3-3-1 model with
right-handed neutrinos" in which the leptons $(\nu_l\; l^-\,
\nu^c_l)^T_L$ transform as $(\textbf{3},-1/3)$. If there are not
right-handed neutrinos in nature the later model should be ruled
out. On the other hand, if these neutrinos do really exist it
suggests that the larger symmetry among neutral and singly charged
leptons could be $SU(4)_L\otimes U(1)_N$, transforming like
$(\nu_l\; l^-\, \nu^c_l\, e^+)^T_L\sim (\textbf{4},0)$~\cite{su4}.
There exist other models which include leptons that are not of the
known lepton species but include heavy neutrinos (they have
right-handed singlets associated to them) $(\nu_l\; l^-\,
\nu^\prime_l)^T_L$~\cite{singer} or heavy charged leptons $(\nu_l\;
l^-\, E^+_l)^T_L$~\cite{pt} and $(\nu_l\; l^-\,
E^-_l)^T_L$~\cite{ozer}. Some of these models have a more economic
scalar sector and could serve as a laboratory to explore ideas and
mechanism in the context of a 3-3-1 gauge symmetry. For instance, if
in a given model with only three triplets, as in the model of
Ref.~\cite{pt}, it is possible to implement soft CP violation
through three complex VEVs and a complex trilinear term in the
scalar potential~\cite{cp6}, then it is certain that that mechanism
of CP violation will also work in the minimal model which has four
complex VEVs and two complex trilinear coupling constants (the
opposite is not necessarily true)~\cite{331}. We would like to
stress that some models do not have the same SM weak isospin
attribution~\cite{pt2} and can be phenomenologically ruled out.
4)~It is interesting that the minimal model can be embedded in a
Pati-Salam-like model with $SU(4)_{PS}\otimes SU(4)_{L+R}$ gauge
symmetry~\cite{ssen}, where the $SU(3)_L$ subgroup which contains
the vector bosons of the SM should be the minimal 3-3-1 model of
Sec~\ref{subsec:mm}. This may indicate the route toward a grand
unification theory of three family 3-3-1 models.

\acknowledgments

A. G. D. is supported by FAPESP under the process 05/52006-6, and
this work was partially supported by CNPq under the processes
305185/03-9 (J.C.M.) and 306087/88-0 (V.P.).

\appendix

\section{Global approximate $SU(3)_{L+R}$ symmetry}
\label{sec:a1}

Let us consider the model showed in Sec.~\ref{subsec:hl} which has
the minimal scalar content: only the three triplets $\eta$, $\rho$
and $\chi$ given in Eq.~(\ref{trim1}).  With them we can define the
3-triplet
\begin{equation}
\Phi=\frac{1}{\sqrt3}(\rho\;\chi\;\eta).
\label{33}
\end{equation}

The gauge-covariant derivative is
\begin{equation}
D_\mu\Phi=\partial_\mu\Phi +ig\mathcal{M}_\mu\Phi +ig_X\, B_\mu
\Phi\hat{X}, \label{c1}
\end{equation}
where $\hat{X}=\textrm{diag}(+1,-1,0)$ and the matrix
$\mathcal{M}_\mu$ is defined in Eq.~(\ref{mbv}) with $b=\sqrt3$.

The scalar Lagrangian is written as,
\begin{eqnarray}
{\cal L}(\Phi)&=&\textrm{Tr}[(D_\mu\Phi)^\dagger
D^\mu\Phi]-\mu^2\textrm{Tr}(\Phi^\dagger\Phi)-
\bar{\lambda}_1[\textrm{Tr}(\Phi^\dagger\Phi)]^2-\bar{\lambda}_2
\textrm{Tr}(\Phi^\dagger\Phi)^2 \nonumber \\ &-& \frac{\sqrt3}{2}\,f
\epsilon_{ijk}\epsilon_{mnl}\Phi_{im} \Phi_{jn}\Phi_{kl}+H.c.,
\label{c2}
\end{eqnarray}
which is invariant under global and local $SU(3)_L\otimes U(1)_X$ gauge transformations:
respectively
\begin{equation}
\Phi\to L\Phi,\quad \Phi\to \Phi e^{i\hat{X}\theta}.
\label{c3}
\end{equation}

Since it is the $U(1)_X$ charge that distinguishes the triplets
$\eta,\rho$ and $\chi$, in the limit $g_X=0$ (which is the same that
$\sin\theta_W=0$ by Eq.~(\ref{polo})), the scalar Lagrangian has an
additional $SU(3)_R$ global symmetry under which we have
\begin{equation}
\Phi\to \Phi R^\dagger,
 \label{c4}
\end{equation}
and we see that in this limit the Higgs sector of the 3-3-1 model
has an accidental global symmetry:
\begin{equation}
SU(3)_L\otimes SU(3)_R,\quad \Phi\to L\Phi R^\dagger,
\label{c5}
\end{equation}
where $SU(3)_L$ is the global version of $SU(3)_L$ gauge symmetry,
and $SU(3)_R$ is an approximate accidental global symmetry. This
symmetry implies in Eq.~(\ref{potencialsu2})
$\mu^2_\eta=\mu^2_\rho=\mu^2_\chi\equiv \mu^2/3$, and relations
between $\lambda_1,\cdots, \lambda_{10}$ in Eq.~(\ref{potencialsu2})
and $\bar{\lambda}_1,\bar{\lambda}_2$ in Eq.~(\ref{c2}).

When the Higgs fields acquire vacuum expectation values we have
\begin{equation}
\langle\Phi\rangle=\frac{1}{\sqrt6}\,\left( \begin{array}{ccc}
v_\rho & 0 &0 \\
0 &v_\chi &0 \\
0 & 0 &v_\eta
\end{array}\right),
\label{c6}
\end{equation}
breaking in this way both $SU(3)_L$ and $SU(3)_R$. When the
condition in Eq.~(\ref{sol3b}) and $s_W=0$ are used
$v_\eta=v_\rho=v_W/\sqrt2$ and we have again the $SU(2)_{L+R}$
global symmetry considered in Sec.~\ref{sec:custodial}.

We have verified that the scalar
potential defined in Eq.~(\ref{c2}) has the appropriate number of Goldstone bosons
and the mass spectra of all the charged and neutral sectors have realistic values.
For instance, the physical singly charged
scalar fields have square masses given by
\begin{eqnarray}
M^2_{1^+}=\frac{1}{18}\left[\bar{\lambda}_2(v^2_\eta+v^2_\rho)-\frac{9f}{\sqrt2}
\left(\frac{v_\eta}{v_\rho}+\frac{v_\rho}{v_\eta} \right)v_\chi \right],\nonumber \\
M^2_{2^+}=\frac{1}{18}\left[\bar{\lambda}_2(v^2_\eta+v^2_\chi)-\frac{9f}{\sqrt2}
\left(\frac{v_\eta}{v_\chi}+\frac{v_\chi}{v_\eta} \right)v_\rho \right],
\label{ec1}
\end{eqnarray}
while the double charged scalar has a square mass
\begin{equation}
M^2_{++}=\frac{1}{18}\left[ \bar{\lambda}_2 (v^2_\rho+v^2_\chi)-\frac{9f}{\sqrt2}
\left(\frac{v_\rho}{v_\chi}+\frac{v_\chi}{v_\rho} \right)v_\eta \right].
\label{ec2}
\end{equation}
In the pseudoscalar sector we have
\begin{equation}
M^2_A=-\frac{f}{\sqrt2}\left(\frac{v_\eta v_\rho}{v_\chi}
+\frac{v_\eta v_\chi}{v_\rho}
+\frac{v_\rho v_\chi}{v_\eta} \right),
\label{eni}
\end{equation}
which implies that $f<0$. The real neutral Higgs scalars have square masses which
depend on $\bar{\lambda}_1$ and $\bar{\lambda}_2$ but we will not write them here,
given only numerical results.

The constraint equations from the stationary condition of the scalar
potential are:
\begin{eqnarray}
3\mu^2+(\bar{\lambda}_1+\bar{\lambda}_2)v^2_\eta+\bar{\lambda}_1(v^2_\rho+v^2_\chi)+
\frac{9}{\sqrt2}\,\frac{fv_\rho v_\chi}{v_\eta}=0\quad (a),
\nonumber \\
3\mu^2+(\bar{\lambda}_1+\bar{\lambda}_2)v^2_\rho+\bar{\lambda}_1(v^2_\eta+v^2_\chi)+
\frac{9}{\sqrt2}\,\frac{fv_\eta v_\chi}{v_\rho}=0\quad(b),
\nonumber \\
3\mu^2+(\bar{\lambda}_1+\bar{\lambda}_2)v^2_\chi+\bar{\lambda}_1(v^2_\eta+v^2_\rho)+
\frac{9}{\sqrt2}\,\frac{fv_\eta v_\rho}{v_\chi}=0,\quad(c).
\label{vinculos}
\end{eqnarray}

These equations should be solved for the parameters
$\bar{\lambda}_1,\bar{\lambda}_2$ and $f$, in terms of $\mu^2$ and
the VEVs $v_\eta,v_\rho$ and $v_\chi$. If the VEVs are left free we
get $\bar{\lambda}_1=-3\mu^2/ (v^2_\eta+v^2_\rho+v^2_\chi)$,
$\bar{\lambda}_2=0$ and $f=0$. Since this is not a realistic
scenario, we will impose some constraints on the VEVs. One of the
more interesting possibility that we have found is: assuming
$\bar{\lambda}_2=0$ and $v_\eta=v_\rho=v_\chi\equiv v_W/\sqrt2$, in
this $\langle\Phi\rangle$ in Eq.~(\ref{c6}) implies a global
$SU(3)_{L+R}$ symmetry.   It also implies
$\bar{\lambda}_1=-3\sqrt{2}f/(v_W-4\mu^2/v_W)$ and, using $f=-120$
GeV and $\mu=80i$ GeV implies $\lambda_1=2.49$ (which is within the
perturbative regime). With these input parameters we get that all
the charged scalar masses are of the order of 102 GeV. The
pseudoscalar $A$ has a mass of the order of 144 GeV and in the real
neutral scalar sector we obtain an scalar with mass of the order of
121 GeV and two others mass degenerate states with 144 GeV. This is
just an illustration, the important point is the fact that with the
potential (\ref{c2}) it is possible to obtain realistic values for
the physical scalar masses.

The approximate $SU(3)_L\otimes SU(3)_R$ symmetry occurs in the
Yukawa sector as well. We can see this by defining, using the quark
representation in Eqs.~(\ref{q1l}) and (\ref{q1r}), the 3-triplet
$F=(f_1\,f_2\,f_3)/\sqrt3$ with
\begin{equation}
f_1=Y^\prime_{3\alpha}\overline{Q}_{3L}\,d_{\alpha R},\quad \quad
f_2=Y\overline{Q}_{3L}\,J_R,\quad
f_3=Y^{\prime\prime}_{3\alpha}\overline{Q}_{3L}\,u_{\alpha R},
 \label{f}
\end{equation}
which transform under $SU(3)_C\otimes SU(3)_L\otimes U(1)_X$ as
$(\textbf{1},\textbf{3}^*,-1), (\textbf{1},\textbf{3}^*,+1)$ and
$(\textbf{1},\textbf{3}^*,0)$, respectively, and $Y,Y^\prime$ and
$Y^{\prime\prime}$ are Yukawa couplings. Similarly we define another
3-triplet $G=(g_1\,g_2\,g_3)/\sqrt3$ with
\begin{equation}
g_1=G_{m\alpha}\overline{Q}_{mL}\,u_{\alpha R},\quad
g_2=G^\prime_{m\alpha}\overline{Q}_{mL}\,j_{mR}, \quad
g_3=G^{\prime\prime}_{m\alpha}\overline{Q}_{mL}\,d_{\alpha R},
 \label{g}
\end{equation}
transforming as
$(\textbf{1},\textbf{3},+1),(\textbf{1},\textbf{3},-1)$ and
$(\textbf{1},\textbf{3},0) $, respectively. With (\ref{f}),
(\ref{g}) and the 3-triplet $\Phi$ defined in Eq.~(\ref{33}) we can
re-write the usual Yukawa couplings in the following way
\begin{equation}
-\mathcal{L}=\textrm{Tr}(F\Phi^T)+\textrm{Tr}(G\Phi^\dagger),
 \label{yukaq}
\end{equation}
which is manifestly $SU(3)_L\otimes SU(3)_R$ invariant if $G$ and
$F$ transform as $\Phi$ in Eq.~(\ref{c5}) in the limit $g_X=0$.
Moreover, if we want that the custodial be also manifest in the
lepton sector we see that it is mandatory to add right-handed
neutrinos. We can then define the 3-triplet, using the leptons in
Eqs.~(\ref{lep2l}) and (\ref{lep2r}),
$\psi_{ab}=(\overline{\Psi}_{aL} l_{bR}\;
\overline{\Psi}_{aL}E_{bR}\;\overline{\Psi}_{a}\nu_{bR})/\sqrt3$, so
that the Yukawa interactions can be written as
\begin{equation}
-\mathcal{L}=h_{ab}\textrm{Tr}(\psi_{ab}\Phi^T).
 \label{yukal}
\end{equation}
Notice that the Dirac mass of the neutrinos are equal to the masses
of the charged leptons. This demand the introduction of Majorana
mass terms for the right-handed components, which are singlets of
the $SU(3)_{L+R}$, in such a way that we can implement the see-saw
mechanism for generating small neutrino masses. This also happens in
the SM: it is necessary to add right-handed neutrinos if we want to
implement a custodial symmetry in the lepton sector by defining the
2-doublet $D_{ab}=(\overline{L}_al_{bR}\;\overline{L}_a\nu_{bR})$.
Thus, we can write the Yukawa coupling as
$-\mathcal{L}=h_{ab}\textrm{Tr}(D_{ab}\varphi)$, where
$\varphi=(H,\tilde{H})$ with $H$ the usual Higgs scalar doublet and
$\tilde{H}=\epsilon H^*$.

In the model with the sextet, the sextet is just a symmetrized
3-triplet and we can write down the scalar potential invariant under
$SU(3)_L\otimes SU(3)_R$ as in Eq.~(\ref{c2}), using $\Phi$ defined
in Eq.~(\ref{33}) and the sextet $S$. However, notice that, if
$\langle \sigma^0_1\rangle\not=0$ this breaks also the $SU(2)_{L+R}$
global symmetry imposing a strong constraint on this
VEV~\cite{comment}. The model with right-handed neutrinos may be
considered in the same way.

\end{document}